\documentclass[final,5p,times,twocolumn,authoryear]{elsarticle}

\usepackage{framed,multirow}
\usepackage{amssymb}
\usepackage{latexsym}
\usepackage{url}
\usepackage{hyperref}
\usepackage{xcolor}
\usepackage{amsmath,amsfonts,xfrac}
\usepackage[normalem]{ulem}
\DeclareMathOperator*{\argmax}{argmax}

\definecolor{newcolor}{rgb}{.8, .349, .1}

\graphicspath{
	{./artwork/}
}

\usepackage{numcompress}\biboptions{authoryear}

\begin{document}


\makeatletter
\def\ps@pprintTitle{%
	\let\@oddhead\@empty
	\let\@evenhead\@empty
	\let\@oddfoot\@empty
	\let\@evenfoot\@oddfoot
}
\makeatother
\begin{frontmatter}

\title{Automatic Semantic Segmentation of the Lumbar Spine: Clinical Applicability in a Multi-parametric and Multi-centre Study on Magnetic Resonance Images} 


\author[jj]{\includegraphics[scale=0.05]{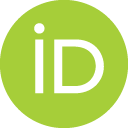}
	\hspace{-2px}\href{https://orcid.org/0000-0003-3332-9710}{
		Jhon Jairo S\'{a}enz-Gamboa}\corref{cor1}}	
\cortext[cor1]{Corresponding authors:  
	jhonjsaenzg@gmail.com (J.J. S\'{a}enz-Gamboa), 
	delaiglesia\_mar@gva.es (M. Iglesia-Vay\'{a})}

\author[jd]{\includegraphics[scale=0.05]{orcid.png}
	\hspace{-2px}\href{https://orcid.org/0000-0001-9488-8159}{
		Julio Domenech}}

\author[AA]{Antonio Alonso-Manjarr\'{e}s}

\author[ja]{\includegraphics[scale=0.05]{orcid.png}
	\hspace{-2px}\href{https://orcid.org/0000-0002-4174-3762}{
		Jon A. G\'{o}mez}}

\author[jj,miv]{\includegraphics[scale=0.05]{orcid.png}
	\hspace{-2px}\href{https://orcid.org/0000-0003-4505-8399}{
		Maria de la Iglesia-Vay\'{a}}\corref{cor1}}

\address[jj]{\textit{FISABIO-CIPF Joint Research Unit in Biomedical Imaging,
		Fundaci\`{o} per al Foment de la Investigaci\`{o} Sanit\`{a}ria
		i Biom\`{e}dica (FISABIO),
		Av. de Catalunya 21,
		46020 Val\`{e}ncia, Spain}}
\address[jd]{\textit{Orthopedic Surgery Department,
		Hospital Arnau de Vilanova,
		Carrer de San Clemente s/n,
		46015, Val\`{e}ncia, Spain}}
\address[AA]{\textit{Radiology Department,
		Hospital Arnau de Vilanova,
		Carrer de San Clemente s/n,
		46015, Val\`{e}ncia, Spain}}
\address[ja]{\textit{Pattern Recognition and Human Language Technology research center,
		Universitat Polit\`{e}cnica de Val\`{e}ncia,
		Cam\'{i} de Vera, s/n,
		46022, Val\`{e}ncia, Spain}}
\address[miv]{\textit{Regional ministry of Universal Health and Public Health in Valencia,
		Carrer de Misser Masc\'{o} 31,
		46010 Val\`{e}ncia, Spain}}

\begin{abstract}
	One of the major difficulties in medical image segmentation is the high variability of these images, 
	which is caused by their origin (multi-centre), the acquisition protocols (multi-parametric), 
	as well as the variability of human anatomy, the severity of the illness, 
	the effect of age and gender, among others.
	The problem addressed in this work is the automatic semantic segmentation 
	of lumbar spine Magnetic Resonance images using convolutional neural networks.
	The purpose is to assign a class label to each pixel of an image. 
	Classes were defined by radiologists and correspond to different structural elements 
	like vertebrae, intervertebral discs, nerves, blood vessels, and other tissues.
	The proposed network topologies are variants of the U-Net architecture. 
	Several complementary blocks were used to define the variants: 
	Three types of convolutional blocks, spatial attention models, deep supervision and multilevel feature extractor.
	This document describes the topologies and analyses the results of the neural network 
	designs that obtained the most accurate segmentations. 
	Several of the proposed designs outperform the standard U-Net used as baseline, 
	especially when used in ensembles where the output of multiple neural networks is 
	combined according to different strategies.
\end{abstract}

\begin{keyword}
	Magnetic Resonance Images\sep Spine\sep Semantic Image Segmentation\sep Convolutional Neural Networks\sep 
	Deep Learning\sep Ensembles of Classifiers
		\MSC[2020] 92B20 \sep 92C50 \sep68T07 \sep 68T45 \sep 68U10 \sep 92B10
\end{keyword}

\end{frontmatter}


\section{Introduction}
\label{sect:intro}

Magnetic Resonance (MR) images are obtained by means of a technique based on magnetic fields where
frequencies in the range of radio waves (8--130 MHz) are used.
%
%
This technique obtains medical images with the maximum level of detail so far.
In recent years, MR images became essential to obtain quality images from any part of the human body
thanks to the fact that 
MR images provide either functional and morphological information of both anatomy and pathological processes;
with a spatial resolution and constrast much higher than the obtained by means of other techniques
for medical image acquisition.
Concerning lumbar pathologies, MR imaging is the preferred type of images between radiologists and physicians
specialized in the lumbar spine and the spine in general.
Thanks to MR images they can find disorders in nerve structures, vertebrae, intervertebral discs, muscles and
ligaments with much more precision than ever \citep{roudsari2010lumbar}.

Manual inspection and analysis carried out by human experts (typically radiologists) is the most common
methodology to extract information from MR images.
Visual inspection is carried out slide by slide in order to determine the location, size and pattern of
multiple clinical findings in the lumbar structures, that can be either normal or pathological.
Manual inspection of slides has a strong dependency on the experience of each expert,
so that the variability due to different criteria of experts is a challenge that cannot be ignored
\citep{carrino2009lumbar, berg2012reliability}.
Radiologists, even those with a great experience, need a lot of time to perform the visual inspection
of images, so this is a very slow task as well as tedious and repetitive.
In fact, the excess of information to be processed visually causes fatigue and loss of attention, which leads 
radiologists to not perceive some obvious nuances because of the ``temporary blindness due to workload excess''
\citep{konstantinou2012visual}.

Current progress of Artificial Intelligence (AI) and its application to medical imaging is providing new and more
sophisticated algorithms based on Machine Learning (ML) techniques.
These new algorithms are complementary to the existing ones in some cases, but in general they perform much better
because most of the existing ones are knowledge based (do not learn from data).
The new algorithms are much more robust to detect the lumbar structures
(i.e., vertebrae, intervertebral discs, nerves, blood vessels, muscles and other tissues)
and represent a significant reduction in the workload of radiologists and traumatologists
\citep{coulon2002quantification,van2005semi,de2014robust,de2015automatic}.

In the context of AI, automatic semantic segmentation 
is currently the most widely used technique \citep{litjens2017survey}.
This technique classifies each individual pixel from an image into one of several classes or categories;
each class or category corresponds to a type of objects from real world to be detected.
In recent years, Convolutional Neural Networks (CNNs) are considered the best ML technique to address semantic segmentation tasks.
However, CNNs require a very large amount of manually annotated images to properly estimate the values of the millions
of weights corresponding to all the layers of any CNN topology designed by a Deep Learning (DL) expert.
Robustness and precision of any classifier based on CNNs strongly depend on the number of samples available to train
the weights of the CNN. So, the challenge in all the projects addressing the task of semantic segmentation is the
availability of large enough datasets of medical images.
In order to have a minimum of samples to train models, a manual segmentation procedure was designed in this work,
where both MR image types T1w and T2w were used to manually adjust the boundaries between structural elements and tissues.
Subsection \ref{image:labels:ground:truth} provides more detail about both MR image types.

The main objective of this study is to use a limited dataset of MR images to reach an accurate and efficient segmentation
of the structures and tissues from the lumbar region by means of using individually optimized CNNs or ensembles of several CNNs;
all the used topologies were based on the original U-Net architecture, i.e., they are variants from the U-Net.

\medskip
This paper is organised as follows:
Section \ref{sect:sota} reviews the state of the art and references other works 
related to the automatic semantic segmentation of medical images.
Section \ref{sect:resources} provides details about the used resources, 
where
Subsection \ref{sect:dataset} describes the dataset used in this work, and
Subsection \ref{sect:soft:and:hw} provides details of the hardware infrastructure and software toolkits.
Section \ref{sect:methodology} describes the block types used in this work to design
CNN topologies as variants from the original U-Net architecture.
Section \ref{sect:experiments} describes the experiments carried out in this work.
Sections \ref{sect:results} and \ref{sect:discussion} present and discuss the results respectively.
Finally, Section \ref{sect:conclusions} concludes by taking into account the defined objectives
and draws possible future works.

\section{Related work}
\label{sect:sota}

Fully Convolutional Networks (FCNs) are one of the topologies of Deep Neural Networks (DNNs) successfully used
for semantic sementation \citep{long2015fully}.
FCNs come from the adaptation of CNNs used for image classification, and generates a map of spatial labels as output.
FCNs are compared with AlexNet \citep{krizhevsky2017imagenet}, VGG16 \citep{simonyan2014very} and
GoogLeNet \citep{szegedy2015going} in \cite{long2015fully}.
The topology known as FCN-8 that comes from an adaptation of VGG16 was the one which obtained the best results 
on the 2012 PASCAL VOC segmentation challenge \citep{everingham2010pascal}.

Notwithstanding, FCNs present an important limitation to cope with semantic segmentation:
the fixed size of the receptive field cannot work with objects whose size is different,
the effect is that such objects are fragmented or missclassified.
Furthermore, relevant details of the objects are lost because the deconvolution process
is too coarse \citep{noh2015learning}.

New approaches arose to overpass limitations of FCNs.
A subset of the new approaches come from the FCNs and use a deep deconvolution. 
Both
SegNet \citep{badrinarayanan2015segnet, badrinarayanan2017segnet} and DeConvnet \citep{noh2015learning} 
belong to this subset.
SegNet is an autoencoder based on convolutional layers, where each layer in the encoder branch is
paired with a layer in the decoder branch, in the sense their shapes are the same.
The \emph{softmax} activation function is used at the output of the last layer of the decoder branch.
The addition of deeper encoder-decoder layer pairs provides a greater spatial context,
what leads to smoother predictions and better accuracy as more pairs are added.
The potential in performace of SegNet is shown in \cite{al2019boundary}, where it is proposed a methodology to detect lumbar
spinal stenosis in axial MR images by means of semantic segmentation combined with boundary delimitation.

The network architecture that is currently obtaining the best results is the U-Net \citep{ronneberger2015u}.
This is a encoder-decoder architecture whose main feature is layer mergence by 
concatenating the features of the layers at the same level of depth,
these concatenations are known as skip connections. 
U-Net has been used with success for the semantic segmentaion in medical images of
liver \citep{christ2016automatic},
kidney \citep{cciccek20163d},
skin lesions \citep{lin2017skin},
prostate \citep{yu2017volumetric}, 
retinal blood vessels \citep{xiao2018weighted},
eye iris \citep{lian2018attention} and
brain structures \citep{roy2018quicknat}
,
and especially in spine \citep{friska_2018_development, sudirman_2019_lumbar, huang_2020_spine, li_2021_automatic}.

%
%

%
This work is an extension of our previous one focused on the task of segmenting MR sagittal
images to delineate structural elements of the anatomy of the lumbar region
\citep{saenz2020semsegspinal}.
There, we analysed some variations of the U-Net architecture by using
(a) convolutional blocks \citep{simonyan2014very, ronneberger2015u},
(b) spatial attention models \citep{schlemper2019attention},
(c) deep supervision \citep{zeng20173d, goubran2020hippocampal} and
(d) multi-kernels at input \citep{szegedy2015going};
the last one is based on a naive version of the architecture Inception \citep{szegedy2015going}.
The integration of these block types improved the performance of the original U-Net architecture.
However, not all the topologies, designed by combining different block types, obtained good results
due to the limited size of the dataset available when the experimentation was carried out.
In our previous work we used manually annotated MR slides from $75$ patients,
in this work we used slides from $181$ patients.

In order to improve the results obtained by classifiers when operating alone,
a widely used strategy is the use of ensembles of classifiers,
that is,
combinations of predictive models with similar but different features.
In an ensemble, the predictions of several classifiers are combined to reduce the variance,
assuming that the type of errors of one classifier is different from the others \citep{goodfellow2016deep}.
In general, the prediction accuracy of an ensemble is better than the accuracy
of each single classifier used in the ensemble \citep{bishop1995neural}.

A comparative study of the performance of four strategies to combine the output of classifiers within ensembles
for image recognition tasks is presented in \cite{ju2018relative}.
The four strategies are ``Unweighted Average'' \citep{breiman2001random}, ``Majority Voting'',
``Bayes Optimal Classifier'' and ``Stacked Generalization'' \citep{wolpert1992stacked,van2007super}.
The study presents experiments in which distinct network structures with different control points were used, 
and analyses the problem of overfitting, a typical problem of neural networks, and its impact on ensembles.
Other approaches using ensembles in semantic segmentation tasks are based on transfer learning, where
networks trained with different datasets from the one of the target task are retrained \citep{nigam2018ensemble},
or are based on ``Stacked U-Nets'' trained in two stages.
In this last case, ensembles of classifiers are used to detect morphological changes in the cell nucleus
from the automatic segmentation of both nuclei regions and regions of overlapping nuclei \citep{kong2020nuclear}.
The relevance of ensembles leads to work in which model compression techniques are applied to
achieve real-time performance to do predictions in production environments \citep{holliday2017speedup}.

%

In this work, we propose new network topologies derived from the U-Net architecture
which are improvements of the topologies we presented in our previous work \cite{saenz2020semsegspinal}.
The results presented here 
were obtained using both individual networks and ensembles.
The proposed ensembles combine distinct network topologies.
The dataset used to obtain the results presented here is an extension of the one used in our
previous work where more manually segmented MR images from other patients have been added.

\section{Resources}
\label{sect:resources}

\begin{figure*}[t]
	\centerline{\includegraphics[]{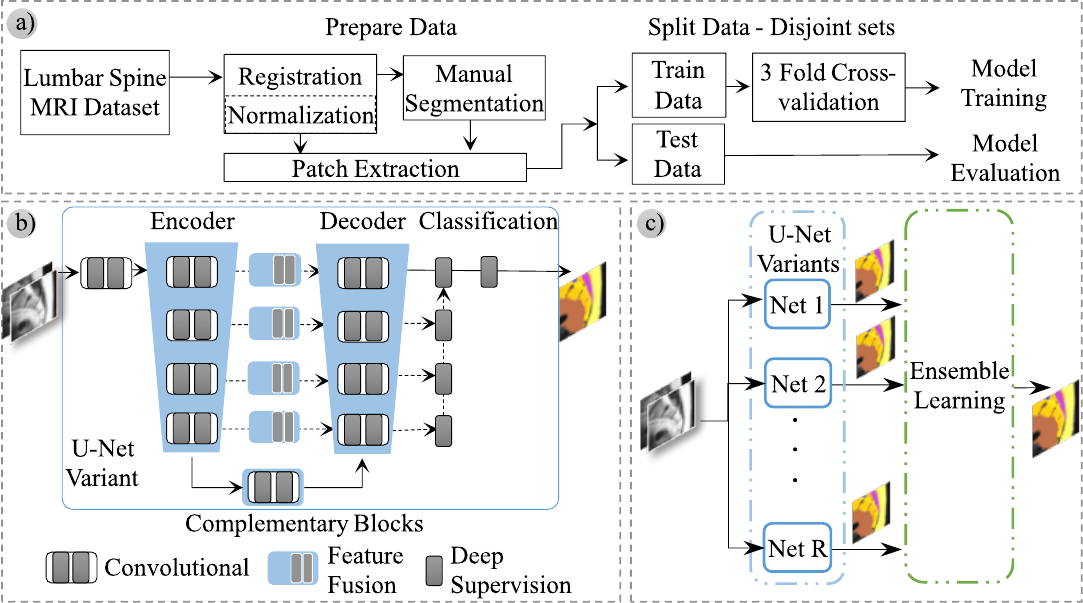}}
	\caption{Steps taken in this work: 
		a) data preparation and manual segmentation to create the ground-truth metadata,
		b) design of the modular framework to easily define U-Net variants, and
		c) evaluation of individual networks and ensembles to create more sophisticated
		models by combining different topologies.}
	\label{fig:modularArch}
\end{figure*}

Figure \ref{fig:modularArch} schematically shows the sequence of steps followed in this work. 
In the first step, the lumbar spine MR imaging dataset was selected, processed and
partitioned into two subsets, one for training and validating corresponding to $80\%$ of patients,
and another for testing with images from the remaining $20\%$ of the patients.
In turn, the first subset was partitioned into two subsets:
one of them to train the models ($53\%$ of the entire dataset and referred to as the training subset)
and the other to adjust hyperparameters according to the results obtained ($27\%$ of the entire dataset
and referred to as the validation subset).
This way of partitioning the largest subset was repeated three times in order to obtain
three pairs of training and validation subsets to evaluate all the models in a 
$3$-fold cross-validation procedure.

In the second step, it was designed the modular framework from which distinct network topologies
derived from the U-Net architecture can be easily defined;
each derived topology was the result of combining several complementary and interchangeable blocks. 
Finally, the design and evaluation of distinct 
topologies
was carried out in the third and last step,
where different configurations of ensembles were also evaluated.

It can be observed that all the variants derived from the U-net architecture have two branches.
The descending branch plays the role of an encoder whereas the ascending one acts as a decoder.
Both branches have four levels in all the variants tested in this work,
and are connected by a bottleneck block in the deepest level.
The classification block is connected to the top layer of the decoder branch and includes the output layer.
%
Predictions from the best variants were combined using Ensemble Learning techniques
\citep{perrone1992networks,bishop1995neural,goodfellow2016deep}.
Results of both individual networks and ensembles are presented in Section \ref{sect:results}, and
the different ensembling strategies are detailed in Subsection \ref{subsect:ensembles}.


\subsection{Lumbar Spine MR Imaging Dataset}
\label{sect:dataset}

The MIDAS dataset is a large collection of Magnetic Resonance (MR) images corresponding
to the lumbar spine. This dataset is one of the main outcomes of the homonym project
Massive Image Data Anatomy of the Spine (MIDAS).
All the images from the same scanning session are accompanied with the report generated by the radiologist who performed the scan.
In numbers, the MIDAS dataset contains more than 23,600 studies with a total of more than 124,800 MR images.
All the studies and images correspond to patients who presented lumbar pathologies
during 2015 and 2016, and were attended in the Health Public System of the Valencian Region.
The public use of the MIDAS dataset was approved by the Ethics committee DGSP-CSISP Nº 20190503/12
once all the data (images, DICOM metadata and reports from radiologists) were properly 
anonymised by the 
\emph{``Banco de Im\'{a}genes M\'{e}dicas de la Comunidad Valenciana''} (BIMCV) \citep{de2014bimcv}
(\url{https://bimcv.cipf.es/bimcv-projects/project-midas/}).
Data management and organisation, including the process of data curation, was done by
following the standard Medical Imaging Data Structure (MIDS) \citep{saborit2020medical}. 

The dataset used in this work is a subset of the MIDAS dataset, where all the selected images
were converted from DICOM format to NIfTi, and the reports, together with other metadata,
were stored using the JSON format.
The hierarchical organization of the NIfTI and JSON files follows the same tree structure of
MIDS, where all the images of a particular scan are located in the same directory, and the
directories of all the sessions belonging to one patient are in the same directory
of a higher level.

\subsubsection{Selection and preparation of the dataset}
\label{sect:dataset:preprocessing}

\begin{table}
	\begin{center}
		\caption{\label{table:demographic}Demographic Statistics of the 181 patients whose scans were used in this work}
		\begin{tabular}{|l|r|r|r|r|} 
			\hline
			& \textbf{Mean} & \textbf{Std} & \textbf{Min} & \textbf{Max} \\ 
			\hline
			Age (year)  & $53$            & $16.5$         & $9$            & $88$         \\ 
			\hline
			Weight (kg) & $74.1$          & $14.6$         & $29$           & $120$        \\ 
			\hline
		\end{tabular}
	\end{center}
\end{table}

The ground-truth for the task of semantic segmentation was generated by manually segmenting a subset of
the MIDAS dataset obtained by randomly selecting studies corresponding to 181 patients.
Each study contains several scanning sessions and each session several MR images.
The age of selected patients ranges from 9 to 88 years, with an average of 53 years, and an unbalanced
gender distribution with 105 women and 76 men.
Table \ref{table:demographic} provides some statistics of the dataset used in this work to carry out all
the experiments.
The studies used in this work were selected according to the following criteria:
\begin{itemize}
	\item Lumbar vertebrae must be included, together with other adjacent anatomical elements, in particular the upper sacral bones.
	\item Each scan should have available both types of sagittal MR images (T1w and T2w) because will be jointly used as input to the models.
	\item 
	T1w and T2w
	from each study must fulfil with
	a predefined quality requirements in terms of brightness and noise.
	\item Selected patients cannot have lumbar surgery.
\end{itemize}
%


Due to the different scan devices used (distinct manufacturers and different models),
the MR images were acquired with different settings parameters,
but the magnetic field intensity was of $1.5$ Teslas in all cases.
Table \ref{tab:scan:settings} lists the range of values for the relevant
configuration parameters according to the metadata accompanying each MR image.

\begin{table}
	\begin{center}
		\caption{\label{tab:scan:settings}Ranges of values of the most relevant configuration parameters of the scan devices}
		\begin{tabular}{|l|c|c|}
			\hline
			View   Plane Types & \multicolumn{2}{c|}{Sagittal}     \\ \hline
			Sequence Types     & T1-weighted     & T2-weighted     \\ \hline
			\begin{tabular}[c]{@{}l@{}}Repetition Time \\ ($ms$)\end{tabular} &
			\begin{tabular}[c]{@{}c@{}}300.0 to \\ 764.38\end{tabular} &
			\begin{tabular}[c]{@{}c@{}}2000.0 to \\ 10172.214\end{tabular} \\ \hline
			Echo Time ($ms$)     & 6.824 to 17.424 & 84.544 to 145.0 \\ \hline
			\begin{tabular}[c]{@{}l@{}}Spacing Between \\ Slices ($mm$)\end{tabular} &
			3.6 to 6.0 &
			3.6 to 6.0 \\ \hline
			\begin{tabular}[c]{@{}l@{}}Imaging Frequency \\ ($MHz$)\end{tabular} &
			\begin{tabular}[c]{@{}c@{}}42.568 to \\ 127.745\end{tabular} &
			\begin{tabular}[c]{@{}c@{}}42.568 to \\ 127.745\end{tabular} \\ \hline
			Echo Train Length  & 2.0 to 10.0     & 13.0 to 36.0    \\ \hline
			Flip Angle         & 80.0 to 160.0   & 90.0 to 170.0   \\ \hline
			Height ($px$)          & 320.0 to 800.0  & 320.0 to 1024.0 \\ \hline
			Width ($px$)       & 320.0 to 800.0  & 320.0 to 1024.0 \\ \hline
			Pixel Spacing  ($mm$)       & 0.4688 to 1.0  & 0.3704 to 1.0 \\ \hline
			Echo Number        & 0.0 to 1.0      & 0.0 to 1.0      \\ \hline
		\end{tabular}
	\end{center}
\end{table}

Sagittal T1- and T2-weighted slices of each scanning session were aligned at the pixel level
by using the FLIRT functionality \citep{jenkinson2001global, jenkinson2002improved}
of the FSL toolkit \cite{jenkinson2012fsl}.
%
%
%
The input to the neural networks for each single slice is a 3D tensor of $H \times W \times 2$,
where $H$ and $W$ are the height (rows) and the width (columns) of the image in pixels,
and $2$ is the number of channels.
Channel 0 corresponds to T2-weighted and channel 1 to T1-weighted.
Once aligned, all the pixels of both channels (T1w and T2w) are normalised to zero mean and unit variance.
Normalisation  is carried out for each channel independently.

There are a total of 1,572 MR images in our dataset corresponding to different slices of the lumbar spine area.
Most of the slices have an image resolution of $512 \times 512$ pixels. 
The number of slices per scanning session ranges from 8 to 14.

\subsubsection{Image Labels and Ground-Truth Metadata}
\label{image:labels:ground:truth}

The ground-truth metadata for the task of semantic segmentation in this work
consists of bit masks generated from the manual segmentation carried out by
two radiologists with high expertise in skeletal muscle pathologies.

	The ground-truth masks delineate different structures and tissues seen in a lumbar MR image. 
	The selection of these structures and tissues was carried out by medical consensus, 
	attending to the need of the MIDAS project regarding the study of the population with a prevalence of lumbar pain and 
	which presents the following radiographic findings: 
	disc dehydration, loss of disc height, disc herniation, Modic changes, facet hypertrophy, yellow ligament hypertrophy,
	foraminal stenosis, canal stenosis, spondylolisthesis, atrophy of paravertebral musculature and 
	fatty infiltration in the dorsal muscles; thus obtaining the eleven classes of interest.
%

As mentioned above, input for neural networks is composed by T1- and T2-weighted slices aligned at the pixel level.
Sagittal T2-weighted images are characterised by highlighting fat and water within the tissues,
and are used by radiologists to distinguish the anatomical silhouette of the different structural elements of the lumbar region.
Sagittal T1-weighted images highlight fat tissue,
and are used in cases where radiologists have doubts regarding some anatomical
structures or findings, e.g., spinal cavity, epidural fat or radicular cysts.

\begin{figure*}[t]
	\centerline{\includegraphics[]{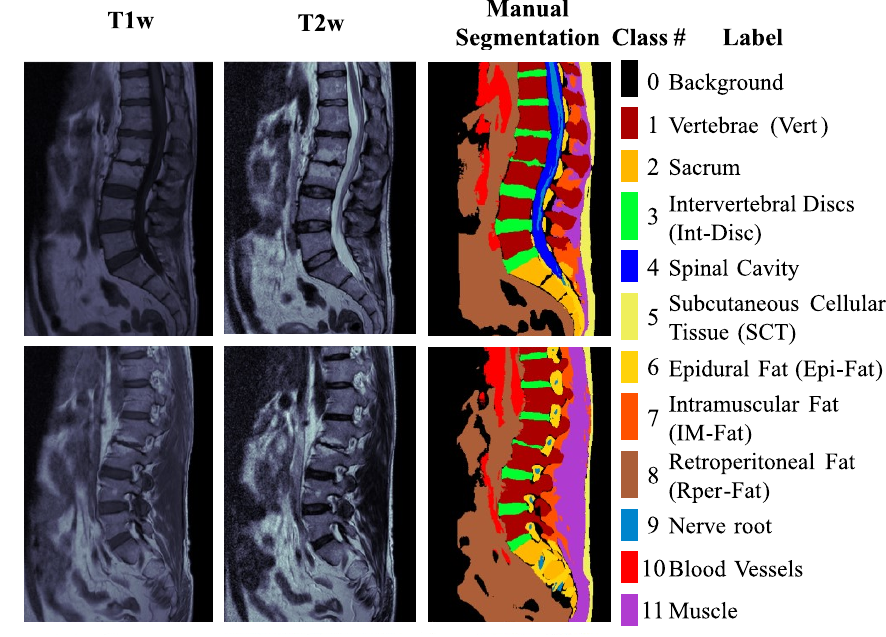}}
	\caption{Example of two different slices with the corresponding bit masks merged
		in a single coloured MR image using one different colour per class.
		From left to right: T1- and T2-weighted MR images, ground-truth semantic segmentation,  
		and labels summary.}
	\label{fig:example:of:image}
\end{figure*}

Figure \ref{fig:example:of:image} shows an example of two different slices
from T1- and T2-weighted sagittal images and their semantic segmentation with
the labels corresponding to 11 target classes plus the background.
The output used to train the neural networks is a stacked 3D tensor containing one bit mask per class.
In other words, the ground-truth masks are tensors of $H \times W \times 12$,
with 12 values per pixel, all set to 0 but one, the value corresponding to the class is set to 1.
Figure \ref{fig:example:of:image} represents each class with a different colour.

%

\subsubsection{Patch Extraction}
\label{subsect:patches}

As indicated in Subsection \ref{sect:dataset}, image acquisition was done with different settings parameters and different sizes.
The dimension of input samples is relevant when using neural networks because height and width in pixels must be fixed at network input.
One possible strategy adopted in many works is to resize all the images to a fixed size.
The strategy followed in this work is different, given an image of $H \times W$ pixels, where both $H$ and $W$ can vary from 320 to 1024,
squared fragments of fixed size $D \times D$ were extracted by shifting approximately $S$ pixels in horizontal and vertical.
Given an input sample, i.e., a 3D tensor with dimensions $H \times W \times 2$,
it is split into overlapping patches with a size of $D \times D \times 2$ extracted using a stride of $S \times S$.
We selected $D = 256$ and $S = 192$ based on experimental results from our previous work \citep{saenz2020semsegspinal}, 
these values for $D$ and $S$ yield a balance between efficiency and accuracy.

\begin{figure*}[t]
	\centerline{\includegraphics[]{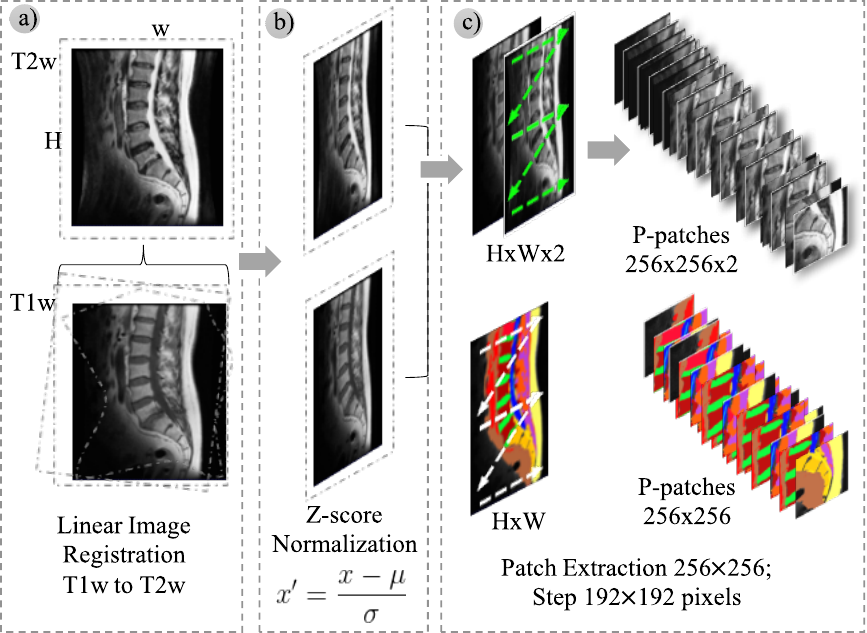}} 
	\caption{Image preprocessing steps:
		(a) linear Image Registration, Sagittal T1-weighted is aligned with T2-weighted,
		(b) both planes (T1- and T2 weighted) are normalised using the Z score procedure,
		(c) joint both 2D slices in a 3D tensor of $H \times W \times 2$, and then,
		(d) split each 3D tensor and its corresponding ground truth mask into overlapping patches of 256 x 256 pixels.}
	\label{fig:preprocessing}
\end{figure*}

In order to prepare training and evaluating samples, the same process of patch extraction
was applied to input images and the corresponding ground-truth masks.
As already mentioned above, ground-truth masks are generated from the manual segmentation.
Table \ref{table:summary} summarises the figures of the dataset, detailing the number of
images per partition, the available 2D slices and the resulting squared fragments or patches.
The set of patients of each partition are disjoint sets, i.e.,
all 2D images (and patches) from one patient are in the same partition.
Figure \ref{fig:preprocessing} shows the image preprocessing steps followed in this work
and the resulting patches as explained in Subsection \ref{sect:dataset:preprocessing}.

\begin{table}
	\begin{center}
		\caption{\label{table:summary}Dataset used for training and testing in figures}
		\begin{tabular}{|l|r|r|r|}
			\hline
			& \textbf{Train \& }  & & \\
			& \textbf{Validation} & \textbf{Test} & \textbf{Total} \\
			\hline
			MR images T2w and T1w         &    $148$  &    $33$  &    $181$ \\ \hline
			Images 2D              &  $1,176$  &   $396$  &  $1,572$ \\ \hline
			Patches 256$\times$256  & $18,147$  & $4,113$  & $22,260$ \\ \hline
		\end{tabular}
	\end{center}
\end{table}

\subsection{Software and Hardware}
\label{sect:soft:and:hw}

The proposed network topologies were implemented 
using TensorFlow \citep{abadi2016tensorflow} and Keras \citep{keras} toolkits.
The linear (affine) image 
transformations were done using FLIRT \citep{jenkinson2001global,jenkinson2002improved}
from the FSL software \citep{jenkinson2012fsl}.
The ground-truth masks were manually segmented by using ITK-SNAP software \citep{py06nimg}.

Training and evaluation was run on the high performance computing infrastructure Artemisa
from the \emph{``Instituto de Física Corpuscular''} \url{https://artemisa.ific.uv.es}
formed by 20 worker nodes equipped with:
2 x Intel(R) Xeon(R) Gold 6248 CPU @ 2.50GHz 20c,
384 GBytes ECC DDR4 at 2933 MHz,
1 x GPU Tesla Volta V100 PCIe.

\section{Methodology}
\label{sect:methodology}

\subsection{Topologies based on the U-Net architecture}
\label{sect:variants}

\begin{figure*}[t]
	\centerline{\includegraphics[]{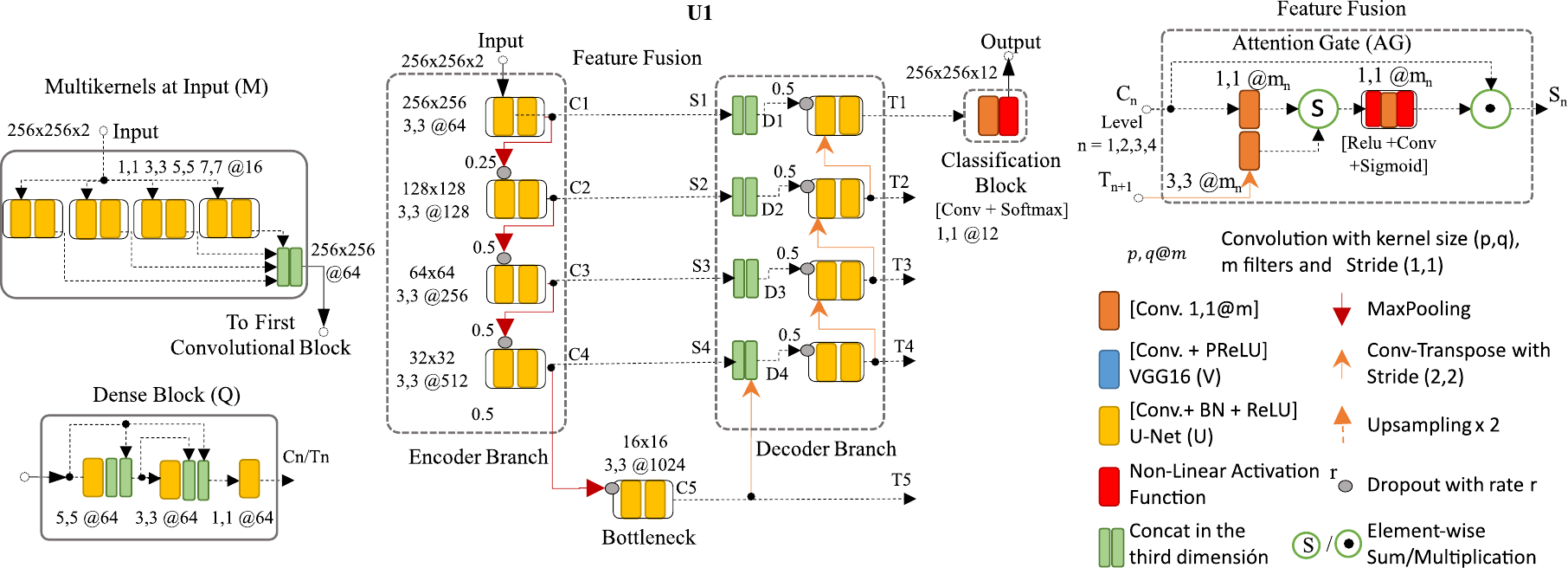}}
	\caption{Example of how the proposed topologies based on the U-Net architecture
		(referred to with the identifier U1 in this document)
		are built from complementary block types:
		(a) Multi-kernels at input (M),
		(b) three types of convolutional blocks (U-net (U), VGG16 (V) and Dense Block (Q)),
		where U and Q are used in both encoder and decoder branches while V is used only in the encoder branch,
		(c) Attention Gates (AG) for replacing the skip connections between encoder and
		decoder branches with the purpose of fusing and selecting
		relevant features at each level between both branches,
		and
		(d) Deep Supervision (DS) illustrated in Figure \ref{fig:ds:blocks}.}
	\label{fig:unet_Blocks}
\end{figure*}

In this work, different topologies have been designed from the U-Net architecture.
Original U-Net architecture is used to obtain baseline results.
To do this, we defined a set of distinct interchangeable block types
which are strategically combined to form encoder and decoder branches.
Some of the topologies presented here were designed using in the decoder branch different block types from the ones used in the encoder branch.
Other topologies use the same block type in both branches.
Figure \ref{fig:unet_Blocks} illustrates an example of a variant from the U-Net architecture
and the distinct block types used in different parts of the topology.
Next subsections are dedicated to explain all the block types used in this work.

\begin{figure*}[h]
	\centerline{\includegraphics[]{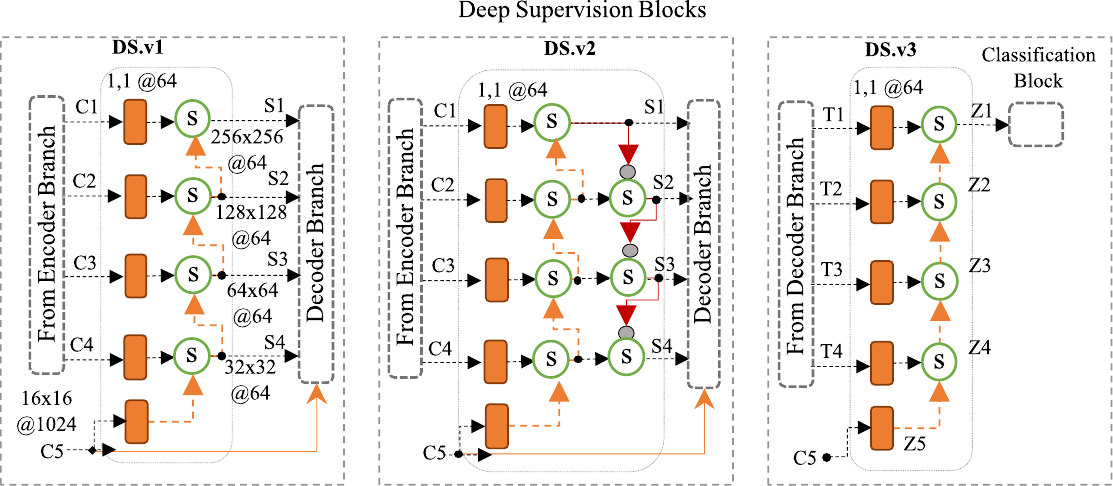}}
	\caption{Deep supervision block types.
		DS.v1 and DS.v2 are used as alternatives to enhance the connections between encoder and decoder branches.
		DS.v3 is used to enrich the input to the classification block; the output of the convolutional block at
		each level of the decoder branch is combined with an element-wise sum with the supervised signals coming
		from the previous level of the decoder branch too.}
	\label{fig:ds:blocks}
\end{figure*}

\subsubsection{Convolutional Block}
Three types of convolutional blocks were tested in this work:
(i) The typical block used in the original U-Net \citep{ronneberger2015u},
which consists of two convolutional layers preceding a batch normalisation
layer that is followed by an activation layer that uses the
``Rectified Linear Unit'' (ReLU).
The size of the kernel for both convolutional layers is $3\times 3$.
(ii) The convolutional block of the VGG16 \citep{simonyan2014very} composed by
two or three convolutional layers with a $3\times 3$ kernel and followed by an
activation layer with ``Parametric Rectified Linear Unit'' (PReLU).
And (iii) the convolutional dense block \citep{roy2018quicknat} consisting of three convolutional layers.
Each convolutional layer of this block type is preceded by a pair of consecutive layers:
a batch normalisation layer followed by an activation layer that uses the ``Rectified Linear Unit'' (ReLU).
The kernel sizes for these three convolutional layers are, respectively, $5\times 5$, $3\times 3$ and $1\times 1$.
The number of channels for the three layers is set to 64.
The input to the second layer is the concatenation of the input to the first layer and the output of the first layer.
The input to the third layer is the concatenation of the input to the first layer, the output of the first layer and the output of the second layer.
\cite{huang2017densely} refers to this type of connections as dense connections.

As Figure \ref{fig:unet_Blocks} shows, the number of filters (or channels) per block
is given by the parameter $m$ at the first (or top) level of the encoder branch
(i.e., the descending path); $m$ is multiplied by 2 when descending from one level to the next one;
except in the case of the convolutional dense block type which was set to 64 for all the levels.
Analogously, $m$ is divided by 2 when ascending from one level to the next one in the decoder
branch (i.e., the ascending path).

\subsubsection{Multi-kernels at Input}
\label{subsubsect:multikernels}
In four of the proposed topologies,
the input layer is connected to a multilevel feature extractor 
rather than using only one convolutional block.
The multilevel feature extractor consists in four independent
convolutional blocks with different kernel sizes
$1\times 1$, $3\times 3$, $5\times 5$ and $7\times 7$.
The output of the four convolutional blocks is concatenated before entering to
the encoder branch in order to extract spatial features at different scales.
This is a variant of the naive version of the Inception network \citep{szegedy2015going}.

\subsubsection{Encoder Branch} 
The encoder branch is made up of four consecutive convolutional blocks.
Each block is followed by a two-dimensional max pooling layer with kernel and stride size equal
to $2\times 2$ to shrink the feature maps to $\sfrac{1}{4}$ in terms of features (rows and columns
divided by 2 each), but maintaining the depth (number of channels).
%

\subsubsection{Feature Fusion} 

Three strategies of Feature Fusion were tested in this work:

\begin{enumerate}[(i)]
	\item The skip connections used in the original U-Net architecture to connect blocks at the
	same level between encoder and decoder branches.
	Feature maps $C_n$ from level $n$ in the encoder branch are concatenated with the
	feature maps $T_{n+1}$ coming from the previous level in the decoder branch.
	This can be seen in Figure \ref{fig:unet_Blocks} where
	$S_n = C_n$ and $D_n = concat(S_n, transposed\_conv(T_{n+1}))$ is the input to the
	convolutional block at level $n$ in the decoder branch.
	The bottleneck output is the special case when $T_5 = C_5$.
	
	\item Deep Supervision (DS).
	
	The underlying idea of Deep Supervision is to provide a complementary feature-map flow in
	the decoder branch.
	We use three versions, DS.v1 and DS.v2 are variants of deep supervision to generate
	complementary input to the convolutional blocks at each level of the decoder branch,
	while DS.v3 takes the outputs from the convolutional blocks of the decoder branch to
	generate a complementary input to the classification block.
	
	Deep supervision was introduced by \cite{lee2015deeply} to perform semantic discrimination
	at different scales in the intermediate layers and also as a strategy to mitigate
	the gradient vanishing problem,
	as shown by \cite{szegedy2015going} in GoogleNet 
	and by \cite{sun2015deepid3} and \cite{shen2019object} in DeepID3.
	
	\medskip
	
	What is proposed in this work as DS.v1 (graphically illustrated in Figure \ref{fig:ds:blocks}) 
	is a deep supervision block to replace the skip connections between encoder and
	decoder branches. Block type DS.v1 is similar to the block used in DeepID3 by
	\cite{sun2015deepid3}, \cite{zeng20173d} and \cite{shen2019object}
	for the same purpose.
	
	In more detail, at each level $n$ of the encoder branch, including the bottleneck,
	the convolutional block generates a feature map, referred to as $C_n$, that is transformed
	by a convolutional layer with a $1\times 1$ kernel with $m$ channels,
	where $m$ is the original number of channels at the first level of the encoder branch.
	%
	%
	The output tensor at the bottleneck level, i.e., the feature map used to start the decoder branch,
	is referred to as $C_5$ in Figure \ref{fig:ds:blocks}.
	The output of the convolutional blocks at each level of the encoder branch are referred to as $C_n$.
	
	When deep supervision is used, all $C_n$ are transformed by a convolutional layer with a $1\times 1$ kernel
	before being combined with the ``supervised signal'' $S_{n+1}$ coming from the previous level.
	In DS.v1, the supervised signals are computed as
	$S_n = conv_{1 \times 1}(C_n) + up\_sampling(S_{n+1})$,
	with the especial case of $S_5 = conv_{1 \times 1}(C_5)$.
	Each $S_n$ is concatenated with $transposed\_conv(T_{n+1})$, i.e.,
	the output of the convolutional block from the previous level in the decoder branch, $T_{n+1}$,
	is transformed by a transposed convolutional before being concatenated with $S_n$
	to obtain the input to the convolutional block at level $n$ of the decoder branch:
	$D_n = concat(S_n, transposed\_conv(T_{n+1}))$,
	as in the case of the original U-Net described above.


	\medskip
	
	A second deep supervision block type, referred to as DS.v2 and also illustrated in Figure \ref{fig:ds:blocks},
	is used between encoder and decoder branches.
	The output of each DS.v2 block at each level is downsampled by a max pooling layer with kernel and stride size
	equal to $2\times 2$, in order to shrink the feature maps to $\sfrac{1}{4}$ in terms of features
	(rows and columns divided by 2 each), while keeping the depth (number of channels) untouched.
	
	The output of a DS.v2 block at one level is combined with the output of the DS.v2 block coming from the above level:
	$S_n = conv_{1 \times 1}(C_n) + max\_pool(S_{n-1})$, with the especial case of the top level in both branches
	where $S_1 = conv_{1 \times 1}(C_1)$.

	\medskip
	
	One additional deep supervision block type referred to as DS.v3 was used to enrich
	the input to the classification block.
	Figure \ref{fig:ds:blocks} illustrates how the output of the convolutional blocks at
	each level of the decoder branch, $T_n$, are combined with the ``supervised signals''
	coming from the previous level, $Z_{n+1}$.
	The supervised signals are upsampled to achieve the same size of $T_n$
	to compute the element-wise sum: $Z_n = conv_{1 \times 1}(T_n) + up\_sampling(Z_{n+1})$,
	being $Z_1$ the input to the classification block in this case.

	DS.v3 block type was also used in our previous work \citep{saenz2020semsegspinal}
	for the same purpose, where was referred to as DS.
	%

	\item Attention Gate (AG).
	In three of the topologies proposed in this work,
	the skip connections between encoder and decoder branches are replaced 
	by a spatial attention model, known as Attention Gate (AG),
	\citep{schlemper2019attention}.
	The purpose of AG is to fuse and select relevant features at each level between both branches.
	Thanks to this, the relevant features automatically selected by the AG from the encoder branch
	are provided to the corresponding level of the decoder branch.
	With this strategy, the different levels of the decoder branch can use the relevant features
	extracted at its paired level in the encoder branch for the progressive reconstruction
	of the output mask.
	AGs only hold relevant features from the encoder branch that are concatenated with the
	feature maps obtained as output of each level in the decoder branch.
	The feature maps from encoder and decoder branches are transformed individually
	by a single convolutional layer with a $1\times 1$ kernel, then are combined with an element-wise
	add operator and passed through a ReLU activation layer followed by another $1\times 1$
	convolutional layer that in turn is followed by a sigmoid activation layer.
	%
	%
	Sigmoid output values within the range [0, 1] act as a 2D mask used to filter
	the feature map coming from the respective level of the encoder branch.
	Then, both the AG output $S_n$ and the feature map from the previous level of the decoder $T_{n+1}$
	are concatenated to connect blocks at the same level;
	as explained previously $D_n = concat(S_n, transposed\_conv(T_{n+1}))$.
	The transposed convolutional resizes $T_{n+1}$ to reach the same size $S_n$ has.
	Transposed convolutional layers are represented in orange arrows in Figure \ref{fig:unet_Blocks}.
\end{enumerate}

\subsubsection{Bottleneck}
The bottleneck is a convolutional block that performs feature estimation 
at an additional depth level and is the main union point between 
encoder and decoder branches.

\subsubsection{Decoder Branch}
The decoder branch consists of a set of four consecutive convolutional blocks,
each one preceded by a feature-fusion block, in such a way that each level of the decoder
branch uses the set of relevant features obtained by fusing both
($a$) the output of the paired convolutional block in the encoder branch with
($b$) the output of the transposed convolutional layer in the previous level of the decoder branch.

Transposed convolutional layers are better at reconstructing the spatial dimension of feature maps
in the decoder branch than performing interpolation using an upsampling layer followed by a normal
convolution.
They can learn a set of weights that can be used to progressively reconstruct original inputs;
but in this work, we use them to generate masks for semantic segmentation.
The use of transposed convolutional layers is very important when the task includes the
segmentation of very small structural elements.

\subsubsection{Classification Block}
The output generated by the last level of the decoder branch,
or the last level of the deep supervision block (DS.v3) when applicable,
is used as input to the classification block.
This block consists in one convolutional layer with a $1\times 1$ kernel and
as many channels as classes into which classify each single pixel.
In our case the number of classes is 12.
The \emph{softmax} activation function was used at the output layer of all the topologies
tested in this work.
In this case, the output values can be considered as \emph{a posteriori} probabilities
normalised over the predicted output classes.
That is, for every pixel of the output mask, each class is weighted by a score in the
range $[0, 1]$ and the sum of the scores of all classes for a single pixel sums $1$.
Accordingly, the ground-truth masks used to train the networks have 12 channels,
in such a way that each single pixel of the output mask is represented by one 1-hot
vector of length 12.
For each pixel of the ground-truth mask only one of the channels is set to 1.

\subsection{Ensembles}
\label{subsect:ensembles}

\begin{figure*}[t]
	\centerline{\includegraphics[]{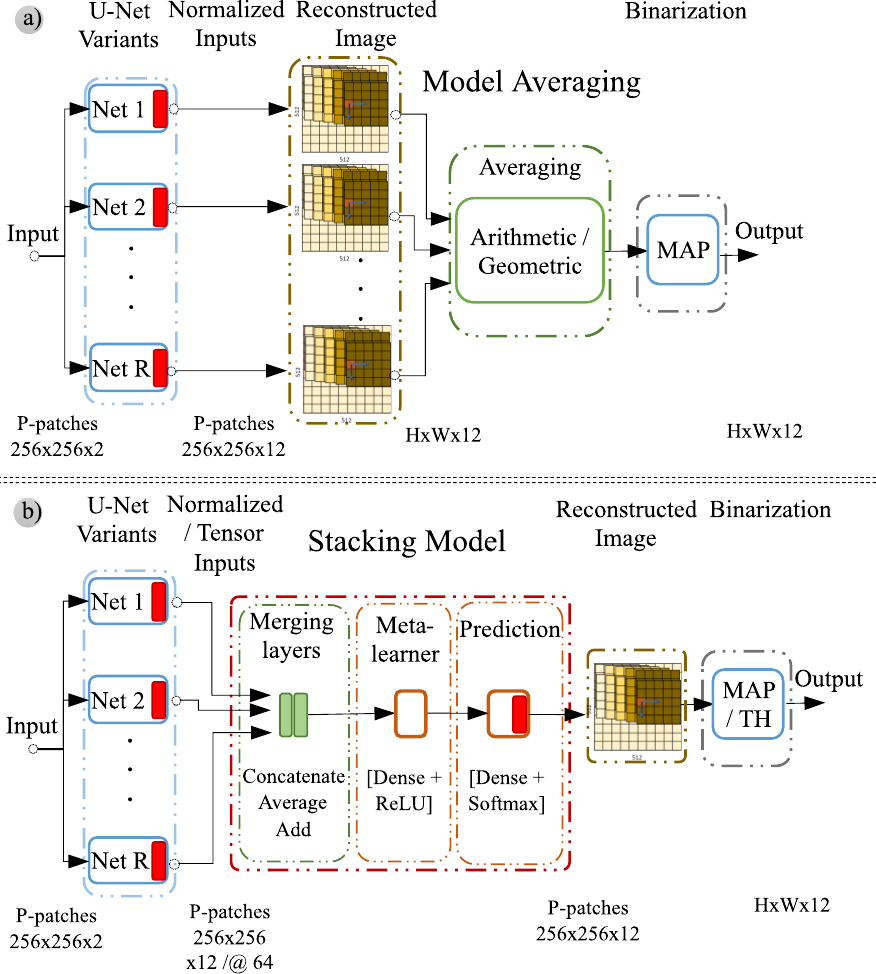}}
	\caption{Block diagram of methods tested in this work to compute the output of ensembles. Top: Model averaging. Bottom: Stacking model.}
	\label{fig:ensembles}
\end{figure*}

In addition to testing with individual networks, every one of the topologies proposed in
this work as variants from the U-Net architecture for the semantic segmentation task was
used in ensembles of several networks. The output of several networks corresponding to
different topologies is combined to form a classifier that is an ensemble of classifiers.
From each topology, it was selected the network that obtained the best results,
i.e., the one adjusted with the best combination of the values of the hyperparameters.
When used in ensembles, the outputs of single classifiers were combined by two distinct
approaches: model averaging and stacking model.
Figure \ref{fig:ensembles} illustrates both approaches.

\subsubsection{Model Averaging}

Model averaging is a technique where $R$ models equally contribute 
to obtain the output of the ensemble, i.e., the prediction provided
by the ensemble is the combination of the prediction of each single model.

%
Two strategies can be used for merging the output of several models: 

\begin{equation} 
	\text{Arithmetic Mean: }
	\overline{Z}={\frac {1}{R}}\sum _{r=1}^{R}Z_{r}
	\label{eqn:arithmetic:c} 
\end{equation}

\begin{equation} 
	\text{Geometric Mean: }
	\overline{Z}={\sqrt[{R}]{\prod _{r=1}^{R}Z_{r}}}
	\label{eqn:geometric:c} 
\end{equation}

\subsubsection{Stacking Model}
\label{subsect:Ensemble:Stacking}

Stacking models learn to obtain a better combination of the predictions of $R$ single models
in order to achieve the best prediction.
An ensemble following the stacking model is implemented in three stages:
(a) \emph{layer merging},
(b) \emph{meta-learner}, and
(c) \emph{prediction}.

The first stage, \emph{layer merging}, takes as input a list of tensors and returns a unique tensor
that could be the result of concatenating, averaging or adding.
The tensors to be merged come from every single model in the ensemble.
They can be the normalised output values, i.e., the output of the \emph{softmax},
or the tensors used as input to the classification block, i.e., the outputs generated
by the last level of the decoder branch or DS.v3 when applicable.
In the second stage a dense layer with a ReLU activation function plays the role of \emph{meta-learner}.
The last stage, \emph{prediction}, consists of a dense layer with the \emph{softmax} activation function.

\subsection{Image Reconstruction and Pixel Level Labelling}
\label{subsect:Binarisation}

The $P$ patches corresponding to an original 2D slice of size $H \times W$,
will be placed in the corresponding position. 
Each pixel of the reconstructed mask can belong to 1, 2 or 4 patches.
In the case of overlapping (i.e., 2 or 4 patches),
the score of each target class per pixel is calculated by using the arithmetic mean of
the occurrences of the respective pixel in the overlapping patches.
Then, each single pixel is labelled with one class according to one of the following two methods.

\subsubsection{Maximum A Posteriori Probability Estimate (MAP)}
The output of the \emph{softmax} activation function in the classification block is a vector of normalised scores,
$y^{m,n} \in \mathbb{R}^{12}$, for each single pixel $X^{m,n}$, where $X$ refers to the input image.
The element $y^{m,n}_{c}$ is the confidence of the network that pixel $X^{m,n}$ belongs to class $c$.
According to the MAP criterion, each single pixel is assigned to the class $c^*$ with the highest score,
i.e., $c^* = \argmax_c\{y^{m,n}_{c}\}$.

\subsubsection{Threshold Optimisation (TH)}

In this work, we used a naive adaptation of the \emph{Threshold Optimisation} strategy
explained in \cite{NIPS2016_Lepora}.
A threshold per target class was tuned using the validation subset of the
three partitions created to carry out the $3$-fold cross-validation procedure.
Section \ref{sect:resources} explains how the dataset was partitioned.
The threshold of each class was adjusted by finding the maximum value of the IoU metric
for different thresholds ranging from 0.05 to 0.95 every 0.05.

Each single pixel at output is assigned to the class with the highest score generated by
the \emph{softmax} activation function if such score is greater than the threshold for such class.
Otherwise, the score of the next best scoring class is checked until
the score of a class that is greater than or equal to its respective threshold is found. 
Classes are checked in descending order according to its score.
The pixel is assigned to the background class if this process ends unsuccessfully.

	MAP or TH will be suffixed to the identifier of each experiment
	to indicate the method used for labelling each single pixel.

\section{Experiments and Implementation}
\label{sect:experiments}

The dataset used in this work was extracted from the MIDAS corpus referenced in Subsection \ref{sect:dataset}.
The MR images come from scanning sessions corresponding to 181 different patients.
Each scanning session has a different number of slices.
How the dataset was partitioned into training, validation and test subsets is explained in Section \ref{sect:resources}.
However, let us emphasize here that all the generated subsets are disjoint at the level of patient, i.e.,
it is guaranteed no 2D images from the same patient appear in different subsets.
Table \ref{table:summary} summarises the figures of the dataset.
The experiments for each evaluated network topology or ensemble were carried out
following the same $3$-fold cross-validation procedure.

As explained in Section \ref{sect:resources},
$80\%$ of patients were used for training and validating and $20\%$ for testing.
In turn, the $80\%$ of patients for training and validating was split into three
different partitions to perform a $3$-fold cross-validation procedure.
In each cross-validation iteration, images from $\sfrac{2}{3}$ and $\sfrac{1}{3}$
of the patients were used for training and validating, respectively.
The reported results were obtained with the test subset as an average of the results obtained
by the three model versions (one per cross-validation iteration).
%

%
%
The reported results were computed after labelling each single pixel with both MAP and TH criteria (see Subsection \ref{subsect:Binarisation}).

\subsection{Data Augmentation}
\label{subsect:DataAugmentation}
In order to mitigate the overfitting problem, training data was randomly modified by the combination
of several 2D image transformations:
($a$) random rotation up to $\pm 20$ degrees,
($b$) zoom in/out by a factor randomly selected from $0.5$ to $1.5$,
($c$) random shift in both axes up to $10\%$ of height and width, and
($d$) horizontal flip according to a Bernoulli probability distribution with $p=\sfrac{1}{2}$.

\subsection{Model hyper-parameters}

\begin{table*}[t]
	\begin{center}
		\caption{\label{table:cnn:settings}Parameter settings of the CNN topologies.
			Network IDs are also used in Table \ref{table:cnn:ensembles} and Table \ref{tab:results:Best:2}.
			DS.v2 is only used in topology UDD2}
			\begin{tabular}{|l|l|l|l|l|}
				\hline
				\textbf{ID} &
				\textbf{Configuration} &
				\textbf{Optimiser} &
				\textbf{Learning Rate} &
				\textbf{Act-Conv} \\ 
				\hline
				UDD2  & U-Net + DS.v3 + DS.v2                       & Adam      & $0.00033$ & ReLU    \\
				UMDD  & U-Net + multi-kernel + DS.v3 + DS.v1        & Adam      & $0.00033$ & ReLU    \\
				UDD   & U-Net + DS.v3 + DS.v1                       & Adam      & $0.00033$ & ReLU    \\
				UQD   & U-Net + DenseBlock + DS.v3                  & Adam      & $0.00033$ & ReLU    \\
				UVDD  & U-Net + VGG16 + DS.v3 + DS.v1               & Adam      & $0.00033$ & PReLU   \\
				UVMD  & U-Net + VGG16 + multi-kernel + DS.v3        & Adam      & $0.00033$ & ReLU    \\
				UAMD  & U-Net + attGate + multi-kernel + DS.v3      & Adam      & $0.00033$ & ReLU    \\
				UMD   & U-Net + multi-kernel + DS.v3                & Adam      & $0.00033$ & ReLU    \\
				UAD   & U-Net + attGate + DS.v3                     & RMSprop   & $0.001$   & ReLU    \\
				UD    & U-Net + DS.v3                               & Adam      & $0.00033$ & ReLU    \\
				UA    & U-Net + attGate                             & Adam      & $0.00033$ & ReLU    \\
				U1    & U-Net                                       & Adadelta  & $1.0$     & ReLU    \\
				FCN   & FCN8                                        & Adam      & $0.00033$ & ReLU    \\
				\hline
			\end{tabular}
	\end{center}
\end{table*}

All the proposed topologies but one are variations from the U-Net architecture.
Let us identify each complementary block with a letter in order to compose the network identifiers:
	\begin{description}
		\setlength{\itemindent}{-16pt}
		\setlength{\labelsep}{8pt}
		\item[A] Attention Gates for replacing the skip connections.
		\item[D] Deep Supervision
		between encoder and decoder branches to replace the skip connections (DS.v1 and DS.v2), and
		between convolutional blocks of the decoder branch to provide an alternative input to the 
		classification block (DS.v3).
		\item[M] A previous step after the input is added just before the first block of the encoder branch.
		This step is defined by several convolutional layers with different kernel sizes whose outputs
		are concatenated (see Subsection \ref{subsubsect:multikernels}). 
		%
		\item[V] Use of VGG16-like convolutional blocks in the encoder branch (i.e. the descending path).
		These convolutional blocks are also connected with the convolutional blocks of the decoder branch.
		\item[U] The typical convolutional block used in the original U-Net.
		\item[Q] Convolutional blocks with dense connections (Dense Block) for replacing U-Net convolutional blocks.  
	\end{description}
%

Table \ref{table:cnn:settings} shows the combination of the configuration parameters that
obtained the best results for each network topology.
All the topologies listed in Table \ref{table:cnn:settings} were trained and evaluated
with different combinations of optimiser, learning rate, and activation function of
the hidden convolutional layers (ReLU or PReLU),
and with same initial number of channels fixed to $64$.
In all cases, the activation function of the output layer was the \emph{softmax} and
the categorical cross entropy was used as the loss function.
%
	Only the results of few topologies and ensembles are reported in this document,
	the results of all the listed topologies are reported in the \nameref{sect:SupplementaryMaterial}.
%
For the sake of brevity, other designed topologies and combinations of configuration parameters
which obtained poor results have been excluded too.
%
%

The two variants that include VGG16 do not use transfer learning,
i.e. the weights of the VGG16 were estimated from scratch. 
In other words, transfer learning is not used in any of the designed and evaluated topologies.
The standard U-Net and the FCN were evaluated to have baseline results.

\subsection{Model training}
All variations designed from the U-Net architecture were trained for 300 epochs
by using the training subset in each of the $3$-fold cross-validation iterations.
The best version of each model at each cross-validation iteration corresponds to
the weight values of the epoch in which the model got the highest accuracy with
the validation subset.

\subsection{Ensembles} 
\label{subsect:Ensembles}

In addition to train and evaluate individual semantic segmentation models designed as
variations from the U-Net architecture, a set of ensembles were created by grouping from 4 to 13 models.
Table \ref{table:cnn:ensembles} shows all the ensembles used in this work, where it can be
observed that the FCN network was only used in ensembles $E8$ and $E13$.
%

\begin{table}[t]
	\begin{center}
		\caption{\label{table:cnn:ensembles}Short names of the ensembles 
			used in this work and the identifiers of the networks
			that constitute each ensemble}
		\begin{tabular}{|l|p{60mm}|}
			\hline
			\textbf{Ensemble Id} & \textbf{Networks (IDs) Included}                               \\ 
			\hline
			$E4$          &              UAD UMD                UQD UDD                             \\ 
			\hline
			$E5$          &           UD UAD UMD UAMD                        UDD2                   \\ 
			\hline
			$E6$          &           UD UAD UMD UAMD UVMD UVDD                                     \\ 
			\hline
			$E7$          &           UD UAD UMD UAMD UVMD      UQD          UDD2                   \\ 
			\hline
			$E8$          & FCN       UD UAD UMD UAMD UVMD      UQD          UDD2                   \\ 
			\hline
			$E9$          &           UD UAD UMD UAMD UVMD UVDD UQD UDD UMDD                        \\ 
			\hline
			$E10$         &           UD UAD UMD UAMD UVMD UVDD UQD UDD UMDD UDD2                   \\ 
			\hline
			$E11$         &        UA UD UAD UMD UAMD UVMD UVDD UQD UDD UMDD UDD2                   \\ 
			\hline
			$E12$         &     U1 UA UD UAD UMD UAMD UVMD UVDD UQD UDD UMDD UDD2                   \\ 
			\hline
			$E13$         & FCN U1 UA UD UAD UMD UAMD UVMD UVDD UQD UDD UMDD UDD2                   \\ 
			\hline
		\end{tabular}
	\end{center}
\end{table}

A dual evaluation was performed to compare the two strategies used in ensembles: model averaging and stacking model.
Additionally, in the case of model averaging, results with the arithmetic \eqref{eqn:arithmetic:c} mean and
the geometric \eqref{eqn:geometric:c} mean were compared.
Figure \ref{fig:ensembles} shows the schemes followed in both model averaging and stacking model techniques.

Let $R$ be the number of models in an ensemble,
let $y_r \in \mathbb{R}^{12}$ be the output of model $r$ for each single pixel with one score $y_{r,c}$ per class (our semantic segmentation task targets $12$ classes),
and let $y \in \mathbb{R}^{12}$ be the output of the ensemble per pixel.
As all the models use the \emph{softmax} activation function in the output layer, their outputs are normalised and sum 1, i.e.,
$\sum_{c} y_{r,c} = 1$
and 
$\sum_{c} y_{c} = 1$.
This is why $y_r$ and $y$ can be considered as vectors of posterior probabilities that we also refer to as vectors of normalised scores.

The model averaging technique computes the score of each class $y_c$ as either the arithmetic mean \eqref{eqn:arithmetic:c} or the geometric mean \eqref{eqn:geometric:c}
from $y_{r,c} \forall r \in [1..R]$.

\begin{table*}[t]
	\begin{center}
		\caption{\label{table:cnn:stacking}Parameter settings of the best performing stacking models}
			\begin{tabular}{|c|c|c|c|c|c|}
				\hline
				\multicolumn{1}{|c|}{\multirow{2}{*}{\textbf{\begin{tabular}[c]{@{}c@{}}Stacking model \\ ID\end{tabular}}}} 
				& \multicolumn{5}{c|}{\textbf{Configuration}} \\ \cline{2-6} 
				\multicolumn{1}{|c|}{} & \textbf{Input} & \textbf{Merging Layers} & \textbf{Meta-learner} & \textbf{Optimiser} & \textbf{Learning Rate} \\ \hline
				NAD                    & Normalised     & Average                 & Dense Layer           & Adam               & 0.00033                \\ \hline
				TCD                    & Tensor         & Concatenate             & Dense Layer           & Adam               & 0.00033                \\ \hline
			\end{tabular}
	\end{center}
\end{table*}

By the other hand, the stacking model technique was used with two different approaches
to prepare the input to the layer-merging stage:
(a) the output of the \emph{softmax} activation layer from each model $r$ in the ensemble, i.e., the vector $y_r$, and
(b) the 64-channel tensor at the input to the classification block, i.e., 
the output generated by the last level of the decoder branch, 
or the last level of the deep supervision block (DS.v3) when applicable.
The combination of the inputs in the layer-merging stage can be done by concatenation, averaging, or adding.

When the inputs to the ensemble are ready, the two dense layers of the stacking model
are trained (see Figure \ref{fig:ensembles}).
The output of the ensemble is also one vector of normalised scores per pixel $y \in \mathbb{R}^{12}$.


Table \ref{table:cnn:stacking} shows input formats and layer configurations for
the best performing ensembles based on the stacking model assembling technique.
Ensemble configurations are identified by a three letter acronynm.
First letter identifies the input type, \textbf{N} and \textbf{T}
which stand for normalised scores (\emph{softmax} output) and 64-channel tensors, respectively.
%
Second letter indicates layer merging operator:
averaging (\textbf{A}) and concatenation (\textbf{C}).
The addition operator was also used in the whole experimentation, however, its results are not presented here because they were too poor.
The third letter corresponds to the type of meta-learner used,
in this case only dense layers were used, so the third letter is fixed to \textbf{D}.

Ensembles based on the stacking model were trained during 50 epochs using the same data-augmentation transformations
used to train each single network (see Subsection \ref{subsect:DataAugmentation}), and following the $3$-fold
cross-validation procedure with the same partitions of the dataset.
The best version of each stacking model at each cross-validation iteration corresponds to the weight values 
of the epoch in which the stacking model got the highest accuracy with the validation subset.

In both assembling strategies, namely, model averaging and stacking model,
the output masks corresponding to $256 \times 256$ patches are used to be combined and generate
a single mask per original slide (medical image) in order to evaluate the quality of the automatic
semantic segmentation.
According to the procedure followed to generate the patches from one slice, each single pixel of
the reconstructed mask can belong to one, two or four patches. In the case of two or four patches
arithmetic mean is used to compute the score of each class within the vector of scores of each single pixel.

The vector corresponding to each single pixel of the reconstructed mask is used to assign each pixel
to one of the 12 classes by either the maximum \emph{a posteriori} probability (MAP) criterion or
the Threshold Optimisation (TH) strategy (see Subsection \ref{subsect:Binarisation}).
Both labelling criteria, MAP and TH, were tested for all single networks and ensembles.

\subsection{Evaluation Metrics}
\emph{Intersection over Union} (IoU) \citep{long2015fully} was used as the metric to compare
the performance of the evaluated network architectures.
IoU is a variant of the Jaccard index to quantify the overlapping between the ground-truth mask
and the predicted mask.
The IoU for each individual class $c$ is defined as follows:
\begin{equation} 
	IoU_{c} = \frac{m_{cc}}{t_{c} + m_{c} - m_{cc}}
	\label{eqn:iou:c} 
\end{equation}
where $m_{cc}$ is the count of pixels of class $c$ correctly predicted by the model into the class $c$,
$t_{c}$ is the total amount of pixels of the class $c$ according to the ground-truth, and
$m_{c}$ is the total amount of pixels assigned to class $c$ by the model.

The global metric reported in the results is the average for all target classes,
i.e., all the classes except the background class.
Averaged IoU is computed according to the following formula:

\begin{equation} 
	IoU = \frac{1}{|C^{*}|} \underset{c \in C^*}\sum IoU_{c}
	\label{eqn:iou} 
\end{equation}
where $C^{*}$ is the set of classes excluding the background class,
i.e., the set of target classes which correspond to each one
of the structural elements to be detected and delimited.

\section{Results}
\label{sect:results}

\if false
\begin{table*}
	\begin{center}
		\caption{\label{tab:results:Best:1}
			Performance of the Automatic Semantic Segmentation generated by several network topologies and ensembles,
			using model averaging in some ensembles and stacking model in others.
			The Intersection over Union (IoU) is the metric used to evaluate the performance with respect to each
			and every one of the 12 classes by using equation \eqref{eqn:iou:c}.
			The average with and without the background class was computed using equation \eqref{eqn:iou}
			-- Background is not a target class}
			\begin{tabular}{|c|l|c|c|c|c|c|c|c|c|c|c|c|}
				\hline
				\multicolumn{13}{|c|}{\textbf{Best performing topologies}} \\ 
				\hline
				\multicolumn{2}{|c|}{\multirow{2}{*}{\textbf{Class}}} & \multicolumn{3}{c|}{\textbf{Baseline}} & \multicolumn{2}{c|}{\textbf{Best Variant}} & \multicolumn{2}{c|}{\textbf{Ensemble Average}} & \multicolumn{4}{c|}{\textbf{Stacking Ensemble}} \\ \cline{3-13} 
				\multicolumn{2}{|c|}{} & \multicolumn{1}{c|}{\multirow{2}{*}{\textbf{\begin{tabular}[c]{@{}c@{}}FCN \\ TH\end{tabular}}}} & \multicolumn{1}{c|}{\multirow{2}{*}{\textbf{\begin{tabular}[c]{@{}c@{}}U1\\ MAP\end{tabular}}}} & \multicolumn{1}{c|}{\multirow{2}{*}{\textbf{\begin{tabular}[c]{@{}c@{}}U1\\ TH\end{tabular}}}} & \multicolumn{1}{c|}{\multirow{2}{*}{\textbf{\begin{tabular}[c]{@{}c@{}}UMD\\ MAP\end{tabular}}}} & \multicolumn{1}{c|}{\multirow{2}{*}{\textbf{\begin{tabular}[c]{@{}c@{}}UMD \\ TH\end{tabular}}}} & \multicolumn{1}{c|}{\multirow{2}{*}{\textbf{\begin{tabular}[c]{@{}c@{}}E13   AA\\ MAP\end{tabular}}}} & \multicolumn{1}{c|}{\multirow{2}{*}{\textbf{\begin{tabular}[c]{@{}c@{}}E13   GA\\ MAP\end{tabular}}}} & \multicolumn{1}{c|}{\multirow{2}{*}{\textbf{\begin{tabular}[c]{@{}c@{}}E10   TCD\\ MAP\end{tabular}}}} & \multicolumn{1}{c|}{\multirow{2}{*}{\textbf{\begin{tabular}[c]{@{}c@{}}E10   TCD\\ TH\end{tabular}}}} & \multicolumn{1}{c|}{\multirow{2}{*}{\textbf{\begin{tabular}[c]{@{}c@{}}E11   NAD\\ MAP\end{tabular}}}} & \multicolumn{1}{c|}{\multirow{2}{*}{\textbf{\begin{tabular}[c]{@{}c@{}}E12   NAD\\ TH\end{tabular}}}} \\ \cline{1-2}
				\multicolumn{1}{|l|}{\textbf{\#}} & \multicolumn{1}{l|}{Id} & \multicolumn{1}{c|}{} & \multicolumn{1}{c|}{} & \multicolumn{1}{c|}{} & \multicolumn{1}{c|}{} & \multicolumn{1}{c|}{} & \multicolumn{1}{c|}{} & \multicolumn{1}{c|}{} & \multicolumn{1}{c|}{} & \multicolumn{1}{c|}{} & \multicolumn{1}{c|}{} & \multicolumn{1}{c|}{} \\ 
				\hline
				0 & \textbf{Background}                & $91,81\%$ & $92,19\%$ & $92,27\%$ & $92,16\%$ & $92,23\%$ & $92,58\%$ & $92,58\%$ & $92,39\%$ & $92,45\%$ & $92,58\%$ & $\boldsymbol{92,61\%}$ \\
				1 & \textbf{Vert}                      & $84,06\%$ & $85,96\%$ & $86,19\%$ & $86,12\%$ & $86,28\%$ & $86,84\%$ & $86,88\%$ & $86,60\%$ & $86,73\%$ & $86,88\%$ & $\boldsymbol{87,01\%}$ \\
				2 & \textbf{Sacrum}                    & $80,99\%$ & $84,09\%$ & $84,30\%$ & $84,38\%$ & $84,75\%$ & $85,24\%$ & $85,27\%$ & $84,83\%$ & $84,99\%$ & $85,08\%$ & $\boldsymbol{85,40\%}$ \\
				3 & \textbf{Int-Disc}                  & $86,91\%$ & $88,69\%$ & $88,87\%$ & $88,86\%$ & $89,10\%$ & $89,35\%$ & $89,36\%$ & $89,13\%$ & $89,29\%$ & $89,43\%$ & $\boldsymbol{89,48\%}$ \\
				4 & \textbf{Spinal-Cavity}             & $72,60\%$ & $75,47\%$ & $75,76\%$ & $75,85\%$ & $76,06\%$ & $76,80\%$ & $76,79\%$ & $76,11\%$ & $76,45\%$ & $76,46\%$ & $\boldsymbol{76,95\%}$ \\
				5 & \textbf{SCT}                       & $91,78\%$ & $92,52\%$ & $92,57\%$ & $92,56\%$ & $92,62\%$ & $93,00\%$ & $93,01\%$ & $92,83\%$ & $92,88\%$ & $93,00\%$ & $\boldsymbol{93,07\%}$ \\
				6 & \textbf{Epi-Fat}                   & $54,60\%$ & $58,00\%$ & $58,31\%$ & $58,51\%$ & $58,87\%$ & $60,00\%$ & $\boldsymbol{60,00\%}$ & $59,12\%$ & $59,44\%$ & $59,57\%$ & $59,95\%$ \\
				7 & \textbf{IM-Fat}                    & $61,07\%$ & $63,78\%$ & $64,02\%$ & $64,23\%$ & $64,56\%$ & $65,48\%$ & $65,49\%$ & $64,76\%$ & $65,05\%$ & $65,43\%$ & $\boldsymbol{65,68\%}$ \\
				8 & \textbf{Rper-Fat}                  & $69,28\%$ & $70,75\%$ & $70,75\%$ & $70,54\%$ & $70,64\%$ & $72,03\%$ & $\boldsymbol{72,04\%}$ & $71,55\%$ & $71,60\%$ & $71,93\%$ & $71,99\%$ \\
				9 & \textbf{Nerve-Root}                & $45,64\%$ & $50,93\%$ & $51,76\%$ & $51,59\%$ & $52,28\%$ & $53,10\%$ & $53,08\%$ & $52,04\%$ & $52,58\%$ & $52,90\%$ & $\boldsymbol{53,27\%}$ \\
				10 & \textbf{Blood-Vessels}            & $58,71\%$ & $60,78\%$ & $61,31\%$ & $60,89\%$ & $61,31\%$ & $63,01\%$ & $62,99\%$ & $62,27\%$ & $62,62\%$ & $63,06\%$ & $\boldsymbol{63,31\%}$ \\
				11 & \textbf{Muscle}                   & $79,40\%$ & $80,82\%$ & $81,08\%$ & $80,96\%$ & $81,22\%$ & $81,90\%$ & $81,91\%$ & $81,44\%$ & $81,62\%$ & $81,92\%$ & $\boldsymbol{82,03\%}$ \\
				\hline
				\multicolumn{2}{|l|}{\textbf{IoU} without Bg.}      & $71,37\%$ & $73,80\%$ & $\boldsymbol{74,08\%}$ & $74,04\%$ & $\boldsymbol{74,33\%}$ & $75,16\%$ & $75,17\%$ & $74,61\%$ & $74,84\%$ & $75,06\%$ & $\boldsymbol{75,29\%}$\\
				\multicolumn{2}{|l|}{\textbf{IoU} with Bg.}        & $73,07\%$ & $75,33\%$ & $75,60\%$ & $75,55\%$ & $75,83\%$ & $76,61\%$ & $76,62\%$ & $76,09\%$ & $76,31\%$ & $76,52\%$ & $\boldsymbol{76,73\%}$\\
				
				\hline
			\end{tabular}
	\end{center}
\end{table*}
\fi

\begin{table*}[t]
	\begin{center}
		\caption{\label{tab:results:Best:2}
			Performance of the Automatic Semantic Segmentation generated by several network topologies and ensembles.
			Some ensembles performed better using model averaging and others using stacking model.
			The Intersection over Union (IoU) was the metric used to evaluate the performance 
			of the 12 classes by using equation \eqref{eqn:iou:c}.
			The average with and without the background class was computed using equation \eqref{eqn:iou}
			-- background is not a target class.
			Ensemble $E13$ obtained good results with both the arithmetic mean and the geometric mean,
			and ensemble $E10$ with both MAP and TH labelling criteria
		}
		\resizebox{1.0\textwidth}{!}{%
			\begin{tabular}{|c|l|c|c|c|c|c|c|c|c|c|c|c|}
				\hline
				\multicolumn{2}{|c}{} & \multicolumn{5}{|c}{} & \multicolumn{6}{|c|}{\textbf{Best performing ensembles}} \\ \cline{3-13}
				\multicolumn{2}{|c|}{}
				& \multicolumn{3}{c|}{\textbf{Baseline}}
				& \multicolumn{2}{c|}{\textbf{Best Variant}}
				& \multicolumn{2}{c|}{\textbf{Model Averaging}}
				& \multicolumn{4}{c|}{\textbf{Stacking Model}} \\ \cline{3-13} 
				\multicolumn{2}{|c|}{}               & FCN & U1  & U1  & UMD & UMD & $E13$ & $E13$ & $E10$ & $E10$ & $E11$ & $E12$ \\
				\multicolumn{2}{|c|}{\textbf{Class}} &     &     &     &     &     & Arith & Geo   & TCD   & TCD   & NAD   & NAD   \\ \cline{1-2} 
				\textbf{\#} & Id                     & TH  & MAP & TH  & MAP & TH  & MAP   & MAP   & MAP   & TH    & MAP   & TH    \\ 
				\hline
				0 & \textbf{Background}                & $91.8\%$ & $92.2\%$ & $92.3\%$ & $92.2\%$ & $92.2\%$ & $92.6\%$              & $\boldsymbol{92.6\%}$ & $92.4\%$ & $92.5\%$ & $\boldsymbol{92.6\%}$ & $\boldsymbol{92.6\%}$ \\
				1 & \textbf{Vert}                      & $84.1\%$ & $86.0\%$ & $86.2\%$ & $86.1\%$ & $86.3\%$ & $86.8\%$              & $86.9\%$              & $86.6\%$ & $86.7\%$ & $86.9\%$              & $\boldsymbol{87.0\%}$ \\
				2 & \textbf{Sacrum}                    & $81.0\%$ & $84.1\%$ & $84.3\%$ & $84.4\%$ & $84.8\%$ & $85.2\%$              & $85.3\%$              & $84.8\%$ & $85.0\%$ & $85.1\%$              & $\boldsymbol{85.4\%}$ \\
				3 & \textbf{Int-Disc}                  & $86.9\%$ & $88.7\%$ & $88.9\%$ & $88.9\%$ & $89.1\%$ & $89.4\%$              & $89.4\%$              & $89.1\%$ & $89.3\%$ & $89.4\%$              & $\boldsymbol{89.5\%}$ \\
				4 & \textbf{Spinal-Cavity}             & $72.6\%$ & $75.5\%$ & $75.8\%$ & $75.9\%$ & $76.1\%$ & $76.8\%$              & $76.8\%$              & $76.1\%$ & $76.5\%$ & $76.5\%$              & $\boldsymbol{77.0\%}$ \\
				5 & \textbf{SCT}                       & $91.8\%$ & $92.5\%$ & $92.6\%$ & $92.6\%$ & $92.6\%$ & $93.0\%$              & $93.0\%$              & $92.8\%$ & $92.9\%$ & $93.0\%$              & $\boldsymbol{93.1\%}$ \\
				6 & \textbf{Epi-Fat}                   & $54.6\%$ & $58.0\%$ & $58.3\%$ & $58.5\%$ & $58.9\%$ & $\boldsymbol{60.0\%}$ & $\boldsymbol{60.0\%}$ & $59.1\%$ & $59.4\%$ & $59.6\%$              & $\boldsymbol{60.0\%}$ \\
				7 & \textbf{IM-Fat}                    & $61.1\%$ & $63.8\%$ & $64.0\%$ & $64.2\%$ & $64.6\%$ & $65.5\%$              & $65.5\%$              & $64.8\%$ & $65.1\%$ & $65.4\%$              & $\boldsymbol{65.7\%}$ \\
				8 & \textbf{Rper-Fat}                  & $69.3\%$ & $70.8\%$ & $70.8\%$ & $70.5\%$ & $70.6\%$ & $\boldsymbol{72.0\%}$ & $\boldsymbol{72.0\%}$ & $71.6\%$ & $71.6\%$ & $71.9\%$              & $\boldsymbol{72.0\%}$ \\
				9 & \textbf{Nerve-Root}                & $45.6\%$ & $50.9\%$ & $51.8\%$ & $51.6\%$ & $52.3\%$ & $53.1\%$              & $53.1\%$              & $52.0\%$ & $52.6\%$ & $52.9\%$              & $\boldsymbol{53.3\%}$ \\
				10 & \textbf{Blood-Vessels}            & $58.7\%$ & $60.8\%$ & $61.3\%$ & $60.9\%$ & $61.3\%$ & $63.0\%$              & $63.0\%$              & $62.3\%$ & $62.6\%$ & $63.1\%$              & $\boldsymbol{63.3\%}$ \\
				11 & \textbf{Muscle}                   & $79.4\%$ & $80.8\%$ & $81.1\%$ & $81.0\%$ & $81.2\%$ & $81.9\%$              & $81.9\%$              & $81.4\%$ & $81.6\%$ & $81.9\%$              & $\boldsymbol{82.0\%}$ \\
				\hline
				\multicolumn{2}{|l|}{\textbf{IoU} without Bg.}     & $71.4\%$ & $73.8\%$ & $74.1\%$ & $74.0\%$ & $74.3\%$ & $75.2\%$ & $75.2\%$ & $74.6\%$ & $74.8\%$ & $75.1\%$ & $\boldsymbol{75.3\%}$\\
				\multicolumn{2}{|l|}{\textbf{IoU} with Bg.}        & $73.1\%$ & $75.3\%$ & $75.6\%$ & $75.6\%$ & $75.8\%$ & $76.6\%$ & $76.6\%$ & $76.1\%$ & $76.3\%$ & $76.5\%$ & $\boldsymbol{76.7\%}$\\
				
				\hline
			\end{tabular}
		}
	\end{center}
\end{table*}

\begin{figure*}[t]
	\centerline{\includegraphics[]{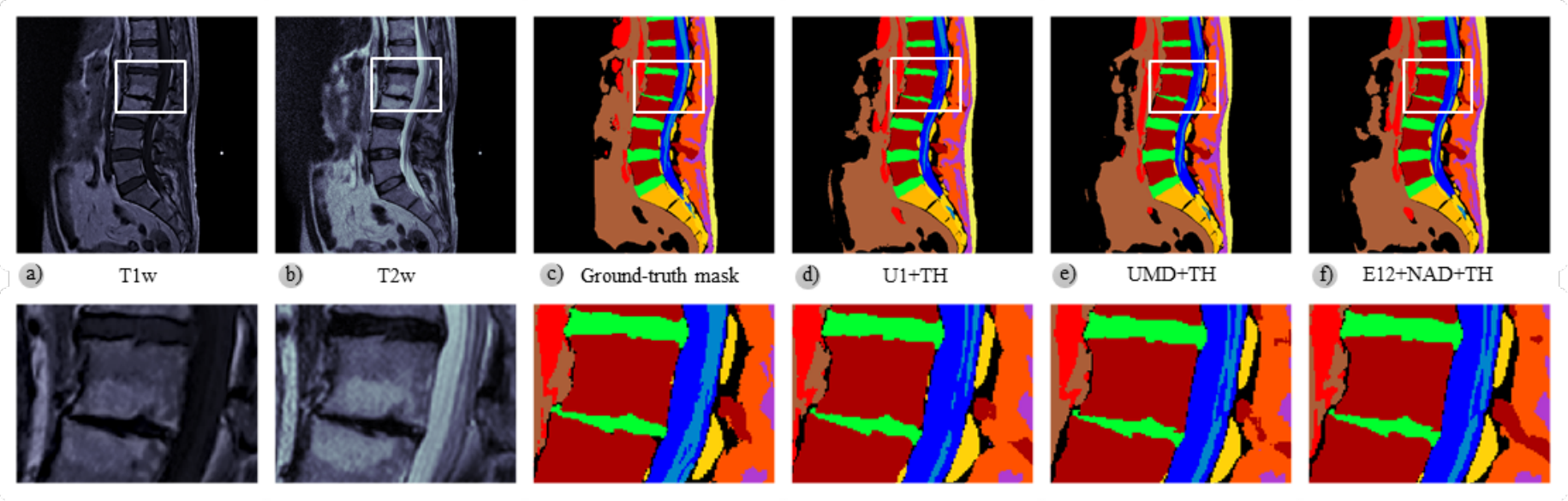}}
	\caption{\label{fig:results:qualitative}Comparison of the qualitative results
		of the best performing topology (UMD+TH) and the best performing ensemble
		($E12$+NAD+TH) with the baseline network architecture (U1+TH).
		A zoomed view shows a posterior protrusion of the L1-L2 disc (green - superior)
		and a marked  L2-L3 disc space narrowing (green - inferior).
		Additionally, the vertebral endplates are affected by Modic changes.
		This example demonstrates the high quality of the semantic segmentation
		obtained despite the variability in morphology and signal of the vertebral
		elements due to the evolution of the pathologies.}
\end{figure*}

The problem addressed in this work is the automatic semantic segmentation of
lumbar spine MR images using CNNs, both single networks and combining the
segmentations generated by several networks within ensembles.
%
%
The goal of the task is to detect and delimit regions in images corresponding to
12 different classes: 11 target classes plus the background.
%


Two criteria described in Subsection \ref{subsect:Binarisation}
were used to label each single pixel into one of the target classes.
The first criterion is based on the \emph{Maximum a Posteriori Probability Estimate} (MAP).
Each single pixel at output is assigned to the class with the highest score generated by the \emph{softmax} activation function.
The second criterion is based on a naive adaptation of \emph{threshold optimisation} (TH).
A threshold per target class was tuned using the validation subset 
to compute the value of the IoU metric for different thresholds. 
The threshold used for each target class was the one which obtained the best performance.

%
%
Let us refresh how topologies presented and evaluated in this work were designed.
Figure \ref{fig:unet_Blocks} shows a diagram of U-Net architecture (U1), used as baseline,
and the complementary blocks used to enhance it. 
All the topologies but the ones used as baseline
were designed as variations from the U-Net by strategically using one or more of the complementary blocks.

%
Table \ref{table:cnn:settings} provides the list of topologies
tested in this work and their respective configuration parameters.
	For the sake of brevity, only the results of few of them are presented
	in this document, those which obtained the highest accuracies.
	In particular, one variant of single networks and four ensembles.
%
The networks architectures U1 and FCN correspond to the standard U-Net
\citep{ronneberger2015u} and FCN8 \citep{long2015fully} architectures.
The results obtained with these two networks were used as the baseline to
compare the results obtained with the proposed variations.


%
Table \ref{table:cnn:ensembles} shows the evaluated ensembles constituted by grouping
different topologies designed as variations from the U-Net architecture.
The listed ensembles are made up from 4 to 13 of the designed network topologies.
The FCN architecture is only used in two ensembles, $E8$ and $E13$, for comparation purposes.

%

Table \ref{tab:results:Best:2} shows the Intersection over Union (IoU) per class computed according
to \eqref{eqn:iou:c} and the averaged IoU calculated according to \eqref{eqn:iou}
for just one topology of single networks, the one which obtained best results, and the four ensembles that performed best.
The results of topologies FCN and U1 are used as the baseline.
The averaged IoU including the background class is only shown for informational purposes.
The best results for each one of the classes have been highlighted in bold.

More specifically, the results of U1, UMD and $E10$ are reported in two columns to show the
effect of the two labelling criteria used in this work, namely MAP and TH.
TH slightly improves the results of MAP in practically all classes, but especially in the case of the
class \emph{Nerve-Root}, precisely the most difficult to be detected.
In the particular case of ensemble $E13$, the two columns show that no differences can be observed between
using the arithmetic mean or the geometric mean; only classes \emph{Vert} and \emph{Sacrum} show a negligible
difference in favor of the geometric mean.
This reflects that all the topologies combined in this ensemble perform very similarly, and,
as it is expected and was previously commented, the use of ensembles lead to more robust and stable
semantic segmentations, what is in line with the observed reduction of the variance of the results
among the cross-validation iterations.
Topology UMD obtained the best results of all the variants tested in this work,
outperforming the baseline architecture U-Net (U1) for all classes using the two 
labelling criteria.
The ensemble $E12$+NAD+TH obtained the best overall results. 
Let us remark that the TH labelling criterion performed better than the MAP criterion
in all the performed experiments.
But as discussed later, differences are not statistically significant.

%
Figure \ref{fig:results:qualitative} illustrates three examples of predicted masks:
one from the best performing topology (UMD+TH)
and another from the best performing ensemble ($E12$+NAD+TH)
that can be compared with the mask of the baseline architecture (U1+TH).
The corresponding T1-weighted and T2-weighted slices used as input to the model
and the ground-truth mask are also shown in Figure \ref{fig:results:qualitative}.

%

%
\begin{figure*}[t]
	\centerline{\includegraphics[]{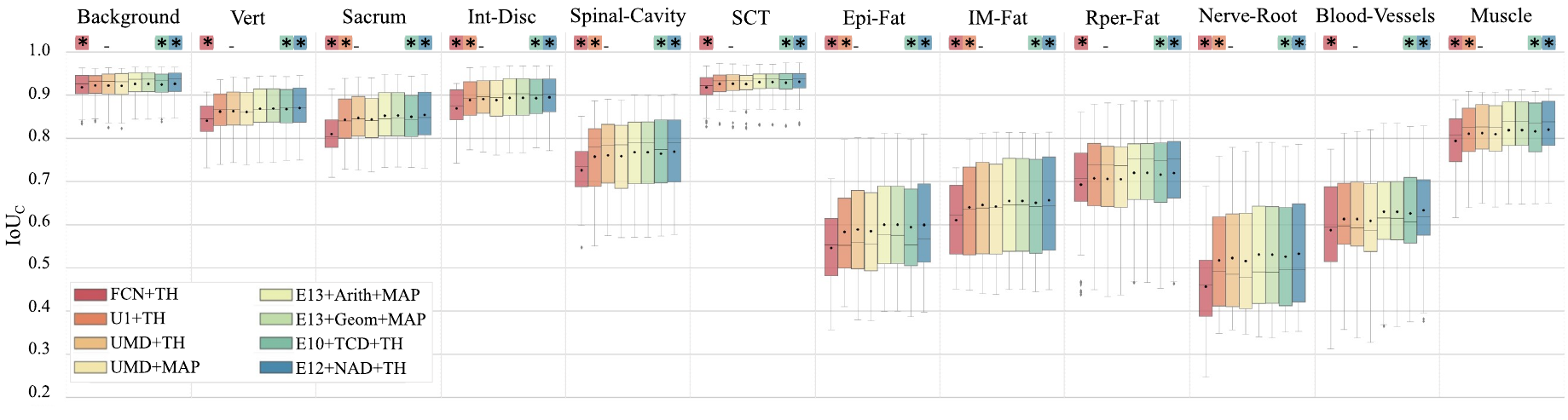}} 
	\caption{\label{fig:box_plot_models}Box plot of Intersection over Union scores per class, $IoU_{c}$,
		for comparing UMD+TH (the best variation from the U-Net architecture) with the best ensembles
		and the two architectures whose results are used as baseline.
		The 11 target structures in the lumbar region plus the background are represented.
		33 MR images from the test subset
		(split into a total of 396 2D overlapping patches of size $256 \times 256$)
		were used for obtaining the classification results to represent the box plots.
		Same classification results were also used 
		for computing the $p$-values according to the Wilcoxon signed-rank test
		in order to check statistical significance of
		model performance differences. 
		Statistical significance ($p < 0.05$) with respect to UMD+TH is indicated by the star symbol ($*$).
	}
\end{figure*}

Figure \ref{fig:box_plot_models} shows the box plot of metric $IoU_{c}$,
i.e., the Intersection over Union score per class,
for comparing the topology derived from the U-Net architecture
that obtained the best results (UMD+TH)
with the best ensembles and the two architectures whose
results were used as baseline.
33 MR images from the test subset
(split into a total of 396 2D overlapping patches of size $256 \times 256$)
were used for obtaining the classification results to represent the box plots.

Additionally, the Wilcoxon signed-rank test was carried out with the same
classification results.
The null hypothesis $H_0$, that in this case can be expessed as
\emph{the mean of the difference of each $IoU_{c}$ is zero},
is not validated in some cases using $0.05$ as the threshold for the $p$-value.
It can be considered that the results of two models are significantly different
when the $p$-value is less than the threshold.
The reference model used for computing the differences was UMD+TH.
Figure \ref{fig:box_plot_models} highlights the models that performed
different with respect to the model UMD+TH according to the Wilcoxon
signed-rank test.
Models are highlighted by means of the star symbol ($*$) and
independently for each target class.

Three observations can be extracted thanks to the Wilcoxon signed-rank test.
First observation, no significant differences in performance between UMD+TH and UMD+MAP exist.
Therefore, as a preliminary conclusion, it can be said that, based on the test subset used in this work,
the labelling criterion TH does not contributes with significant improvements with respect to the MAP criterion.
However, it should be highlighted that the TH labelling criterion depends on
adjusting the threshold of each class by using a different subset than the test subset.
For all the topologies evaluated in this work, the validation subset was used
to adjust the class-dependent thresholds.
It could happen that for other datasets this strategy will not drive to the
optimal thresholds.
Second observation is that UMD+TH performs better than the baseline models.
In seven out of 12 target classes UMD+TH performs better than U1+TH,
and in all target classes UMD+TH outperforms FCN+TH.
Third and last and most important observation is that ensembles
$E10$+TCD+TH and $E12$+NAD+TH performed significantly better than UMD+TH
for all target classes.

\begin{figure*}[t]
	\centerline{\includegraphics[]{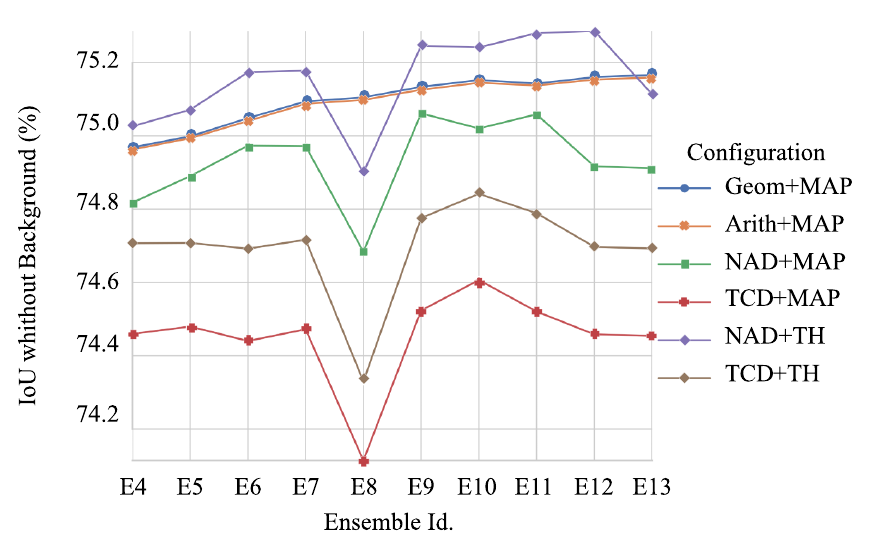}}
	\caption{$IoU$ metric comparing model averaging and stacking model assembling techniques versus the number of networks in each ensemble.}
	\label{fig:comparingEnsembles}
\end{figure*}

Figure \ref{fig:comparingEnsembles} shows
a comparison of the two assembling techniques used in this work, 
model averaging and stacking model.
In the case of model averaging, both ways of computing the output
of the ensemble from the output of the components were considered,
the arithmetic mean and the geometric mean.
In the case of the stacking model technique, two layer merging
strategies are considered: averaging and concatenation.
Averaging uses the vector of normalised scores at the output of the \emph{softmax}.
Concatenation uses the input tensors to the classification block.


First observation from Figure \ref{fig:comparingEnsembles} is that the model averaging
assembling technique is more robust than the stacking model technique
to the variance resulting from the predictions of the networks that constitute the ensemble.
No significant differences between the use of the arithmetic mean and the geometric mean are observed.
As already mentioned above, the high similarity between both ways of computing the mean
confirms that all the topologies combined in the ensembles perform very similarly.

A second observation from Figure \ref{fig:comparingEnsembles} is that
in the case of the stacking model assembling technique, those ensembles including the FCN topology,
$E8$ and $E13$, have a significant performance drop.
Comparing $E12$ and $E13$ results for the configuration NAD+TH, it can be observed that
the addition of the FCN topology significantly deteriorates the performance.

	Due to the fact that it was not possible to assess intra- and inter-observer variability in 
	the manual annotation process, we evaluated, as an alternative, 
	the best-performing topology (UMD+TH) in a similar task.
	The MRI image database we used is publicly available on the Mendeley website \citep{sudirman_2019_lumbar}. There, \cite{friska_2018_development}
	manually labelled axial views of the last three levels of intervertebral discs in 515 scans of subjects with 
	symptomatic back pain. The authors defined the following labels: 
	\emph{intervertebral disc} (IVD), \emph{posterior element} (PE), \emph{thecal sac} (TS)
	and \emph{between the anterior and posterior vertebral elements} (AAP). 
	The authors report high inter-rater agreement in three classes (IVD, PE, TS).
	In another study, \cite{al2019boundary} used the same dataset to segment and detect spinal stenoses, using the U-Net architecture in a network topology they call SegNet-TL80. 
	To compare the results reported by that \cite{al2019boundary}, 
	the UMD+TH topology classification block was adapted to obtain the four classes plus background.
	The model was trained for $30$ additional epochs in the new axial MR imaging context, employing the $3$-fold cross-validation procedure and data augmentation method described above. The remaining $20\%$ of the data was used in the model evaluation.
	
	Table \ref{tab:Exp:SegmExt:Axial:ResultsBest} compares the results reported by \cite{al2019boundary} and those obtained with the UMD+TH topology previously trained on our dataset. 
	As it can be seen, the $UMD$ topology obtained better results, outperforming the reference model Segnet-TL80 in all classes and using the $TH$ labelling criterion.
	

\begin{table}[t]
	\caption{Comparison of the performance of automatic semantic segmentation reported by 
		\textit{SegNet-TL80} \cite{al2019boundary} and generated with \textit{UMD + TH}. 
		The Intersection over Union (IoU) was the metric used to evaluate 
		the performance of the 5 classes in common by using equation \eqref{eqn:iou:c}. 
		The average with and without the background class was computed using equation \eqref{eqn:iou}
		-- background is not a target class.}

	\label{tab:Exp:SegmExt:Axial:ResultsBest}
	\begin{center}
		
		\begin{tabular}{|c|c|c|ll|c|}
			\hline
			\multicolumn{3}{|c|}{\begin{tabular}[c]{@{}c@{}}SegNet-TL80 \\ \cite{al2019boundary}\end{tabular}} & \multicolumn{3}{c|}{UMD + TH} \\ \hline
			\#                & Ax-Label               & IoUc & \multicolumn{1}{c|}{\#} & Sag-Label           & IoUc   \\ \hline
			0                  & BG                     & $98\%$ & \multicolumn{1}{c|}{0}   & Background          & $\boldsymbol{99.4\%}$ \\ \hline
			1                  & IVD                    & $92\%$ & \multicolumn{1}{c|}{4}   & Intervertebral Disc & $\boldsymbol{96.8\%}$ \\ \hline
			2                  & PE                     & $78\%$ & \multicolumn{1}{c|}{1}   & Vertebrae           & $\boldsymbol{91.2\%}$ \\ \hline
			3                  & TS                     & $85\%$ & \multicolumn{1}{c|}{5}   & Spinal Cavity       & $\boldsymbol{90.5\%}$ \\ \hline
			4                  & AAP                    & $53\%$ & \multicolumn{1}{c|}{7}   & Epidural Fat        & $\boldsymbol{74.1\%}$ \\ \hline
			\multicolumn{2}{|c|}{IoU without Bg.} & $77\%$ & \multicolumn{2}{c|}{}                           & $\boldsymbol{88.2\%}$ \\ \cline{1-3} \cline{4-6} 
		\end{tabular}
	\end{center}
\end{table}

In summary, the fact that the variants from the U-Net architecture and thus the
proposed ensembles outperform the proposed baseline in practically all classes
is a fruitful result of this work.
The proposed approach demonstrates high performance in the segmentation of
clinically relevant structures (mainly discs, vertebrae and spinal canal),
despite the variability in the quality and provenance of the MR scans.

\section{Discussion}
\label{sect:discussion}
Data and metadata play a crucial role in this work.
Collecting data was a large and important task that consisted in
(i) centralizing MR images coming from different hospitals with their corresponding
reports generated by radiologists,
(ii) revising the quality of images of each session to decide which
ones are valid to this work, and
(iii) anonymizing both images and reports.
Manually generating the ground-truth mask for each single image was the most
challenging and tedious task.
In fact, as explained in Subsection \ref{sect:dataset} and summarized in
Table \ref{table:summary}, only $1.572$ images from $181$ patients were
manually segmented and used in this work.
The ground-truth masks are the product of the manual semantic segmentation of
images to delimit the 11 target clases plus the background
from the anatomical components of the lumbar region visible in sagittal T1w and T2w MR images.
Each pixel of the ground-truth masks is assigned to one and only one of the 12 classes.
As already mentioned in previous sections of this document,
this work is focused on the lumbar region to automatically delimit anatomical
structures and tissues from sagittal MR images.
Images that come from scanning sessions acquired in different hospitals of the
Valencian region and that can correspond to different pathologies.

%

\subsection{Medical perspective}

In this work, it was designed a specific procedure to semantically segment
structures and tissues of the lumbar region that is based on single CNNs and
ensembles of CNNs.
The procedure performs a multiclass segmentation with promising results
in those structures which are relevant from the clinical point of view:
\emph{vertebrae, intervertebral discs, spinal cavity, muscle, subcutaneous cellular tissue}
and \emph{intra-muscular fat}.

However, the segmentation of other relevant structures like
\emph{nerve root} and \emph{epidural fat} was more difficult 
(nerve roots appear in sagittal slices as very small structures at the level 
of intervertebral foramen).
For these two structures, the highest $IoU_{c}$ obtained with the ensemble $E12$+NAD+TH,
the one that performed best, are $53.3\%$ and $60.0\%$ respectively,
very low values of $IoU$ in comparison with the ones of other structures.

The quality of the segmentation strongly depends on the size of the object to be detected.
In order to mitigate this problem, both intradural and extradural nerve roots are considered as
one single class, the target class \emph{Nerve-root}.
Despite this decision, most errors concerning class \emph{Nerve-root} are false negatives, i.e., 
pixels corresponding to this class are mislabelled as one of the others.

One of the strategies to cope with the problem of the small size of some of the objects to be
detected was the use of multi-kernels, in such a way that the image at the input layer is
processed with receptive fields of different sizes.
The output of the convolutional layers with different kernel sizes
whose input is the input layer is stacked together by concatenation.
Topologies UMD, UMDD, UVMD and UAMD use multi-kernels.
In \cite{jiang2021coronary} they use this multiresolution and multiscale strategy in 
the Coronary vessel segmentation task, 
obtaining promising results against 20 state-of-the-art visual segmentation methods 
using a benchmark X-Ray coronary angiography database.


Analizing other works devoted to the semantic segmentation of brain images \citep{roy2018quicknat},
it can be said that the structural complexity of the lumbar spine is comparable
with the complexity of the brain. There is a high number of structural elements
in both cases, the morphology of which significantly changes between the slices
of the same scanning session.
The number of slices in scanning sessions of the brain is much higher,
in such a way that it is possible to consider all the images from a scan
as a 3D object and rescale it to an isotropic space with a resoluton that
each single pixel of 2D images represents an area of around $1 mm^2$.
It is not possible to do similar transformations using the images available
for this work, because the number of sagittal slices is much lower and not
all scanning sessions have a similar number of slices, i.e., the variance of
the distance between sagittal slices is to high for this purpose.
Additionally, the variations observed in available scannings of the spine,
which are due to ageing and different pathologies, are many more than the ones
observed in the available scannings of the brain.
Usually, patients with different brain and neurological pathologies have much
more similar patterns when compared to patients of distinct spine pathologies.
A good example is the high range of variations due to the degeneration of
intervertebral discs that are a common finding in symptomatic and asymptomatic
individuals \citep{tehranzadeh2000lumbar, lundon2001structure, benoist2005natural}.

\subsection{Limitations}

The following limitations represented important challenges to carrying out all the work described here.

\begin{enumerate}[a)]
	\item MR images were acquired by using distinct models of scanning devices
	and from different manufacturers, that in addition were not calibrated
	exactly the same way.
	Hence, acquisition parameters were not homogeneous.
	In order to minimize the impact of the variability of some of the
	configuration parameters, all the images used in this work were
	selected to ensure that these parameters are within the ranges 
	presented in Table \ref{tab:scan:settings}.
	
	
	Despite the parameter variability, the quality of the automatic
	semantic segmentation confirms the robustness of the models proposed
	in this work,
	and their potential to be used by clinicians.
	
	\item Low image quality due to intrinsic factors of scanning devices like sensibility.
	
	\item 
	Overlapping and ambiguous elements that even medical experts have doubts on to what
	class assign such elements. Therefore, large expertise is required to carry out the
	manual semantic segmentation due to the complexity of the anatomical structures.
	
	The ground-truth metadata was generated by two radiologists.
	Because of the lack of time of the radiologists, the manual segmentation of the
	images from each scanning session was carried out by one of the radiologists.
	Therefore it was not possible to compare different manual segmentations of the
	same images provided by different radiologists.
	On average, one radiologist lasted from five to eight hours to segment the
	12 slices that, on average, come from a scanning session.

	\item The models proposed in this work were not configured to deal with patterns corresponding to
	tissues and findings not included in the training data, as it is the case of tumors and cysts.
	All the elements found during the manual segmentation that do not belong to any of the target
	classes were assigned to the background class.
	
\end{enumerate}

\section{Conclusions and Future Works}
\label{sect:conclusions}

This work addressed the problem of segmenting sagittal MR images corresponding to the lumbar
spine with 11 target classes.
Each target class corresponds to one structural element of the anatomy of the lumbar region.
One additional class, referred to as the background class in this work, was used to help
neural networks to distinguish regions of the image not corresponding to any of the
anatomical structures of interest.
11 different network topologies were designed as variations from the U-Net architecture
to address the problem. These topologies were evaluated both individually and combined
in ensembles.
In light of the results reported in this work, it can be stated the main objective
defined in Section \ref{sect:intro} has been achieved.

Several of the topologies and ensembles of neural networks proposed in this work
outperformed both network architectures, the FCN and the original U-Net,
used as the baseline.

Particularly, the results of the topology UMD and the ensembles E10+TCD+TH and
E12+NAD+TH are significantly better than the results of the baseline architectures
according to the Wilcoxon signed-rank test. Moreover, these two ensembles also
performed significantly better than the topology UMD according to the same
Wilcoxon test.

The use of complementary blocks to enhance the original U-Net architecture improved
its performance. The block types used in this work are deep supervision, spatial
attention using attention gates, multi-kernels at the input and the VGG16 topology
for the encoder branch.
However, the combination of using all the complementary block types did not obtain
the best result. Most variants that included deep super-vision in the decoding
branch improved the baseline.
	Results of all the individual topologies tested are reported in the \nameref{sect:SupplementaryMaterial}.

%
Regarding ensembles, all combinations of topologies trained with the predictions of
individual topologies and following the 3-fold cross-validation procedure with the
same partitions of the dataset performed better than any of the individual
topologies with the validation subset.

The ensembles based on the averaging-model assembling technique showed to be more
robust to the variance of network predictions than the ensembles based on the
stacking-model technique.
In the particular case of the ensembles based on the averaging-model technique,
the results using the geometric mean were slightly better than the obtained using
the arithmetic mean. But the Wilcoxon signed-range test showed that such an
improvement was not statistically significant. However, as mentioned above,
the two ensembles that obtained the best overall results are based on the
stacking model technique.

Intervertebral discs and vertebrae are easier to detect due to the
homogeneity of their textures and their morphology. In future work,
we will concentrate our efforts on the most difficult target classes
to improve the quality of automatic semantic segmentation.
Nerve roots, epidural fat, intramuscular fat and blood vessels are
the most challenging classes due to the heterogeneity of their
morphology and their textures.
Furthermore, nerve roots do not appear in the slices with the same
frequency as other anatomical structures. It is well known that the
imbalance in the number of samples of the different target classes
in the training subset makes the less frequent classes much more
difficult to be detected because the model could not observe enough
samples (2D images in this case) containing regions of such classes.
Imbalance plus heterogeneity of textures and morphologies make it 
especially difficult to detect some classes more accurately.


\section*{CRediT authorship contribution statement}

\textbf{J. J. S\'{a}enz-Gamboa:} Conceptualization, Methodology, Software, Validation, Formal analysis, Investigation, Data curation, Writing - original draft, review, editing \& final version, Visualization.

\textbf{J. Domenech:} Conceptualization, Methodology, Investigation, Resources, Data curation, review. 

\textbf{A. Alonso-Manjarr\'{e}s:} Conceptualization, Methodology, Investigation, Data curation, review. 

\textbf{J. A. G\'{o}mez:} Conceptualization, Methodology, Formal analysis, Investigation, Supervision, Funding acquisition, Writing - original draft, review, editing \& final version.  

\textbf{M. Iglesia-Vay\'{a}:} Conceptualization, Methodology, Formal analysis, Resources, Investigation, Supervision, Funding acquisition, Writing - original draft, review, editing \& final version.

\section*{Acknowledgments}

This work was partially supported by the Regional Ministry of Health of the Valencian Region,
under MIDAS project from BIMCV--\emph{Generalitat Valenciana}, under the grant agreement ACIF/2018/285,
and by
the DeepHealth project,
``Deep-Learning and HPC to Boost Biomedical Applications for Health'',
which has received funding from the European Union's Horizon 2020 research
and innovation programme under grant agreement No 825111.

Authors thanks the Bioinformatics and Biostatistics Unit from Principe Felipe Research Center (CIPF)
for providing access to the cluster co-funded by European Regional Development Funds (FEDER)
in the Valencian Community 2014-2020,
and by the Biomedical Imaging Mixed Unit from
\emph{Fundació per al Foment de la Investigació Sanitaria i biomedica} (FISABIO)
for providing access to the cluster openmind,
co-funded by European Regional Development Funds (FEDER) in Valencian Community 2014-2020.


\section*{Supplementary Materials}
\label{sect:SupplementaryMaterial}

Supplementary material associated with this article can be found, in
the online version, at \url{pending-to-be-assigned}


\bibliography{mybibfile}

\end{document}


	\makeatletter
	\def\ps@pprintTitle{%
		\let\@oddhead\@empty
		\let\@evenhead\@empty
		\let\@oddfoot\@empty
		\let\@evenfoot\@oddfoot
	}
	
	\makeatother

	\begin{frontmatter}
		
		\title{Automatic Semantic Segmentation of the Lumbar Spine: Clinical Applicability in a Multi-parametric and Multi-centre Study on Magnetic Resonance Images \\ Supplementary Material}
		
		\author[jj]{\includegraphics[scale=0.05]{orcid.png}
			\hspace{-2px}\href{https://orcid.org/0000-0003-3332-9710}{
				Jhon Jairo S\'{a}enz-Gamboa}\corref{cor1}}	
		\cortext[cor1]{Corresponding authors:  
			jhonjsaenzg@gmail.com (J.J. S\'{a}enz-Gamboa), 
			delaiglesia\_mar@gva.es (M. Iglesia-Vay\'{a})}

		\author[jd]{\includegraphics[scale=0.05]{orcid.png}
			\hspace{-2px}\href{https://orcid.org/0000-0001-9488-8159}{
				Julio Domenech}}
		
		\author[AA]{Antonio Alonso-Manjarr\'{e}s}
		
		\author[ja]{\includegraphics[scale=0.05]{orcid.png}
			\hspace{-2px}\href{https://orcid.org/0000-0002-4174-3762}{
				Jon A. G\'{o}mez}}
		
		\author[jj,miv]{\includegraphics[scale=0.05]{orcid.png}
			\hspace{-2px}\href{https://orcid.org/0000-0003-4505-8399}{
				Maria de la Iglesia-Vay\'{a}}\corref{cor1}}

		\address[jj]{\textit{FISABIO-CIPF Joint Research Unit in Biomedical Imaging,
				Fundaci\`{o} per al Foment de la Investigaci\`{o} Sanit\`{a}ria
				i Biom\`{e}dica (FISABIO),
				Av. de Catalunya 21,
				46020 Val\`{e}ncia, Spain}}
		\address[jd]{\textit{Orthopedic Surgery Department,
				Hospital Arnau de Vilanova,
				Carrer de San Clemente s/n,
				46015, Val\`{e}ncia, Spain}}
		\address[AA]{\textit{Radiology Department,
				Hospital Arnau de Vilanova,
				Carrer de San Clemente s/n,
				46015, Val\`{e}ncia, Spain}}
		\address[ja]{\textit{Pattern Recognition and Human Language Technology research center,
				Universitat Polit\`{e}cnica de Val\`{e}ncia,
				Cam\'{i} de Vera, s/n,
				46022, Val\`{e}ncia, Spain}}
		\address[miv]{\textit{Regional ministry of Universal Health and Public Health in Valencia,
				Carrer de Misser Masc\'{o} 31,
				46010 Val\`{e}ncia, Spain}}
		
	\end{frontmatter}
	
	
	

	\appendix
	\renewcommand\appendixname{Experiment}
	
	\section{Topologies based on the U-Net architecture}
	\label{sect:Experiment:1}

	In the first experiment, different topologies have been designed from the U-Net architecture. 
	To do this, we defined a set of distinct interchangeable block types which are strategically 
	combined to form encoder and decoder branches (see Subsection 4.1).
	
	Table 4 provides the list of the proposed network architectures are variants of the U-Net and 
	the combination of the configuration parameters that obtained the best results for each network topology, 
	as indicated in Subsection 5.2.
	
	The Lumbar Spine MRI Dataset used in this work was extracted from the MIDAS corpus by randomly selecting 
	studies corresponding to 181 patients (see Subsection 3.1.).
	
	Input for neural networks is composed by T1- and T2-weighted sagittal slices aligned at the pixel level. 
	The ground-truth metadata consists of bit masks generated from the manual segmentation carried out 
	by two expert radiologists with high expertise in skeletal muscle pathology.
	
	All variations designed from the U-Net architecture were trained for 300 epochs by using 
	the training subset in each of the 3-fold cross-validation iterations. 
	The best version of each model at each cross-validation iteration corresponds to the weight values 
	of the epoch in which the model got the highest accuracy with the validation subset.
	
	\medskip
	The reported results were computed after labelling each single pixel to one of the 12 classes 
	with both Maximum a Posteriori Probability Estimate (MAP) and Threshold Optimisation (TH) (see Subsection 4.3).
	
	\begin{table*}[h!] 
		\begin{center}
			
			\caption{Performance of the Automatic Semantic Segmentation generated by several network topologies. 
				The Intersection over Union (IoU) was the metric used to evaluate the performance of 
				the 12 classes by using equation (3). 
				The average with and without the background class was computed equation (4). 
				-- Background is not a target class. 
				The Maximum a Posteriori Probability Estimate (MAP) criteria described in Subsection 4.3
				was used to label each single pixel into one of the target classes.} 
			\renewcommand{\arraystretch}{1.0}
			\resizebox{\textwidth}{!}{%
				\begin{tabular}{|c|l|c|c|c|c|c|c|c|c|c|c|c|c|c|}
					\hline
					\multicolumn{2}{|c|}{\multirow{2}{*}{\textbf{Class}}} & \multicolumn{13}{c|}{\textbf{Labelling according to the MAP criterion.}}                                                                                                                                                                           \\ \cline{3-15} 
					\multicolumn{2}{|c|}{}                                & \multicolumn{2}{c|}{\textbf{Baseline}} & \multicolumn{11}{c|}{\textbf{Variant}}                                                                                                                                                                    \\ \hline
					\textbf{\#} & \textbf{Id} & \textbf{FCN} & \textbf{U1} & \textbf{UA} & \textbf{UD} & \textbf{UAD} & \textbf{UMD} & \textbf{UAMD} & \textbf{UVMD} & \textbf{UVDD} & \textbf{UQD} & \textbf{UDD} & \textbf{UMDD} & \textbf{UDD2}\\
					\hline
					0 & \textbf{Background}                & $91.6\%$ & $92.2\%$ & $92.2\%$ & $92.2\%$ & $92.2\%$ & $92.2\%$ & $92.1\%$ & $92.2\%$ & $92.1\%$ & $92.1\%$ & $\boldsymbol{92.2\%}$ & $92.3\%$ & $92.2\%$ \\
					1 & \textbf{Vert}                      & $83.7\%$ & $86.0\%$ & $85.9\%$ & $\boldsymbol{86.2\%}$ & $86.1\%$ & $86.1\%$ & $86.1\%$ & $86.0\%$ & $86.0\%$ & $85.7\%$ & $86.0\%$ & $86.1\%$ & $86.1\%$ \\
					2 & \textbf{Sacrum}                    & $80.8\%$ & $84.1\%$ & $84.1\%$ & $84.4\%$ & $84.3\%$ & $84.4\%$ & $\boldsymbol{84.4\%}$ & $84.0\%$ & $84.0\%$ & $83.9\%$ & $84.3\%$ & $84.3\%$ & $84.4\%$ \\
					3 & \textbf{Int-Disc}                  & $86.4\%$ & $88.7\%$ & $88.5\%$ & $88.8\%$ & $88.7\%$ & $88.9\%$ & $\boldsymbol{88.9\%}$ & $88.7\%$ & $88.7\%$ & $88.4\%$ & $88.7\%$ & $88.7\%$ & $88.8\%$ \\
					4 & \textbf{Spinal-Cavity}             & $71.9\%$ & $75.5\%$ & $75.7\%$ & $75.6\%$ & $75.6\%$ & $\boldsymbol{75.9\%}$ & $75.8\%$ & $75.4\%$ & $75.6\%$ & $75.5\%$ & $75.6\%$ & $75.6\%$ & $75.6\%$ \\
					5 & \textbf{SCT}                       & $91.7\%$ & $92.5\%$ & $92.5\%$ & $\boldsymbol{92.6\%}$ & $92.6\%$ & $92.6\%$ & $92.5\%$ & $92.3\%$ & $92.4\%$ & $92.3\%$ & $92.5\%$ & $92.5\%$ & $92.5\%$ \\
					6 & \textbf{Epi-Fat}                   & $54.5\%$ & $58.0\%$ & $58.0\%$ & $58.2\%$ & $58.3\%$ & $\boldsymbol{58.5\%}$ & $58.3\%$ & $57.9\%$ & $58.1\%$ & $57.4\%$ & $58.1\%$ & $58.2\%$ & $58.1\%$ \\
					7 & \textbf{IM-Fat}                    & $60.4\%$ & $63.8\%$ & $63.6\%$ & $63.9\%$ & $63.8\%$ & $\boldsymbol{64.2\%}$ & $64.0\%$ & $63.3\%$ & $63.6\%$ & $63.0\%$ & $63.8\%$ & $63.8\%$ & $63.9\%$ \\
					8 & \textbf{Rper-Fat}                  & $69.6\%$ & $70.8\%$ & $70.6\%$ & $\boldsymbol{70.9\%}$ & $70.84\%$ & $70.5\%$ & $70.6\%$ & $70.6\%$ & $70.8\%$ & $70.4\%$ & $70.7\%$ & $70.9\%$ & $70.8\%$ \\
					9 & \textbf{Nerve-Root}                & $44.9\%$ & $50.9\%$ & $51.2\%$ & $51.2\%$ & $51.0\%$ & $\boldsymbol{51.6\%}$ & $51.4\%$ & $51.2\%$ & $50.7\%$ & $50.9\%$ & $51.1\%$ & $51.3\%$ & $51.2\%$ \\
					10 & \textbf{Blood-Vessels}            & $57.8\%$ & $60.8\%$ & $60.4\%$ & $61.1\%$ & $60.8\%$ & $60.9\%$ & $60.9\%$ & $60.4\%$ & $60.8\%$ & $60.3\%$ & $\boldsymbol{61.4\%}$ & $61.1\%$ & $60.7\%$ \\
					11 & \textbf{Muscle}                   & $79.0\%$ & $80.8\%$ & $80.7\%$ & $80.9\%$ & $80.9\%$ & $\boldsymbol{81.0\%}$ & $80.9\%$ & $80.6\%$ & $80.8\%$ & $80.4\%$ & $80.9\%$ & $80.9\%$ & $80.9\%$ \\
					\hline
					\multicolumn{2}{|l|}{\textbf{IoU} without Bg.}      & $71.0\%$ & $73.8\%$ & $73.8\%$ & $74.0\%$ & $73.9\%$ & $\boldsymbol{74.0\%}$ & $74.0\%$ & $73.7\%$ & $73.8\%$ & $73.5\%$ & $73.9\%$ & $73.9\%$ & $73.9\%$ \\
					\multicolumn{2}{|l|}{\textbf{IoU} with Bg.}        & $72.7\%$ & $75.3\%$ & $75.3\%$ & $75.5\%$ & $75.4\%$ & $\boldsymbol{75.6\%}$ & $75.5\%$ & $75.2\%$ & $75.3\%$ & $75.0\%$ & $75.4\%$ & $75.5\%$ & $75.4\%$ \\
					
					\hline
				\end{tabular}
			}
			\label{table:results:MAP}
		\end{center}
	\end{table*}

	\begin{table*}[h!]
		\begin{center}
			\caption{Performance of the Automatic Semantic Segmentation generated by several network topologies. 
				The Intersection over Union (IoU) was the metric used to evaluate the performance of the 12 classes by using equation (3). 
				The average with and without the background class was computed equation (4). 
				-- Background is not a target class. The Threshold Optimisation (TH) criteria described in Subsection 4.3
				was used to label each single pixel into one of the target classes.} 
			\renewcommand{\arraystretch}{1.0}
			\resizebox{\textwidth}{!}{%
				\begin{tabular}{|c|l|c|c|c|c|c|c|c|c|c|c|c|c|c|}
					\hline
					\multicolumn{2}{|c|}{\multirow{2}{*}{\textbf{Class}}} & \multicolumn{13}{c|}{\textbf{Labelling according to the TH criterion.}}                                                                                                                                                                           \\ \cline{3-15} 
					\multicolumn{2}{|c|}{}                                & \multicolumn{2}{c|}{\textbf{Baseline}} & \multicolumn{11}{c|}{\textbf{Variant}}                                                                                                                                                                    \\ \hline
					\textbf{\#} & \textbf{Id} & \textbf{FCN} & \textbf{U1} & \textbf{UA} & \textbf{UD} & \textbf{UAD} & \textbf{UMD} & \textbf{UAMD} & \textbf{UVMD} & \textbf{UVDD} & \textbf{UQD} & \textbf{UDD} & \textbf{UMDD} & \textbf{UDD2}\\
					\hline
					0 & \textbf{Background}                & $91.8\%$ & $92.3\%$ & $92.3\%$ & $92.3\%$ & $92.3\%$ & $92.2\%$ & $92.2\%$ & $92.3\%$ & $92.22\%$ & $92.3\%$ & $92.3\%$ & $\boldsymbol{92.3\%}$ & $92.3\%$ \\
					1 & \textbf{Vert}                      & $84.1\%$ & $86.2\%$ & $86.2\%$ & $86.4\%$ & $86.3\%$ & $86.3\%$ & $\boldsymbol{86.5\%}$ & $86.3\%$ & $86.1\%$ & $86.0\%$ & $86.2\%$ & $86.4\%$ & $86.3\%$ \\
					2 & \textbf{Sacrum}                    & $81.0\%$ & $84.3\%$ & $84.5\%$ & $84.5\%$ & $84.6\%$ & $84.8\%$ & $\boldsymbol{84.8\%}$ & $84.5\%$ & $84.2\%$ & $84.1\%$ & $84.4\%$ & $84.6\%$ & $84.6\%$ \\
					3 & \textbf{Int-Disc}                  & $86.9\%$ & $88.9\%$ & $88.8\%$ & $89.0\%$ & $88.9\%$ & $89.1\%$ & $\boldsymbol{89.1\%}$ & $89.0\%$ & $88.9\%$ & $88.8\%$ & $88.9\%$ & $88.9\%$ & $89.0\%$ \\
					4 & \textbf{Spinal-Cavity}             & $72.6\%$ & $75.8\%$ & $76.1\%$ & $75.9\%$ & $75.8\%$ & $\boldsymbol{76.1\%}$ & $76.1\%$ & $75.8\%$ & $75.9\%$ & $75.7\%$ & $75.9\%$ & $75.9\%$ & $75.9\%$ \\
					5 & \textbf{SCT}                       & $91.8\%$ & $92.6\%$ & $92.6\%$ & $\boldsymbol{92.7\%}$ & $92.6\%$ & $92.6\%$ & $92.5\%$ & $92.4\%$ & $92.5\%$ & $92.4\%$ & $92.6\%$ & $92.6\%$ & $92.5\%$ \\
					6 & \textbf{Epi-Fat}                   & $54.6\%$ & $58.3\%$ & $58.3\%$ & $58.5\%$ & $58.6\%$ & $\boldsymbol{58.9\%}$ & $58.7\%$ & $58.3\%$ & $58.5\%$ & $57.7\%$ & $58.5\%$ & $58.6\%$ & $58.4\%$ \\
					7 & \textbf{IM-Fat}                    & $61.1\%$ & $64.0\%$ & $64.0\%$ & $64.2\%$ & $64.1\%$ & $\boldsymbol{64.6\%}$ & $64.4\%$ & $63.7\%$ & $64.0\%$ & $63.4\%$ & $64.1\%$ & $64.1\%$ & $64.2\%$ \\
					8 & \textbf{Rper-Fat}                  & $69.3\%$ & $70.8\%$ & $70.7\%$ & $\boldsymbol{71.0\%}$ & $70.9\%$ & $70.6\%$ & $70.6\%$ & $70.7\%$ & $70.8\%$ & $70.4\%$ & $70.8\%$ & $71.0\%$ & $70.9\%$ \\
					9 & \textbf{Nerve-Root}                & $45.6\%$ & $51.8\%$ & $51.6\%$ & $51.8\%$ & $51.7\%$ & $\boldsymbol{52.3\%}$ & $52.1\%$ & $51.6\%$ & $51.4\%$ & $51.3\%$ & $51.7\%$ & $52.0\%$ & $51.8\%$ \\
					10 & \textbf{Blood-Vessels}            & $58.7\%$ & $61.3\%$ & $61.0\%$ & $61.4\%$ & $61.3\%$ & $61.31\%$ & $61.3\%$ & $60.9\%$ & $61.3\%$ & $60.9\%$ & $\boldsymbol{61.7\%}$ & $61.4\%$ & $61.2\%$ \\
					11 & \textbf{Muscle}                   & $79.4\%$ & $81.1\%$ & $81.1\%$ & $81.1\%$ & $81.2\%$ & $\boldsymbol{81.2\%}$ & $81.1\%$ & $80.9\%$ & $81.1\%$ & $80.8\%$ & $81.1\%$ & $81.2\%$ & $81.1\%$ \\
					\hline
					\multicolumn{2}{|l|}{\textbf{IoU} without Bg.}      & $71.4\%$ & $74.1\%$ & $74.1\%$ & $74.2\%$ & $74.2\%$ & $\boldsymbol{74.3\%}$ & $74.3\%$ & $74.0\%$ & $74.1\%$ & $73.8\%$ & $74.2\%$ & $74.2\%$ & $74.2\%$\\
					\multicolumn{2}{|l|}{\textbf{IoU} with Bg.}        & $73.1\%$ & $75.6\%$ & $75.6\%$ & $75.7\%$ & $75.7\%$ & $\boldsymbol{75.8\%}$ & $75.8\%$ & $75.5\%$ & $75.6\%$ & $75.3\%$ & $75.7\%$ & $75.7\%$ & $75.7\%$\\
					\hline
				\end{tabular}
			}
			\label{table:result:TH}
		\end{center}
	\end{table*}

	\subsection{Results}
	
	\begin{itemize}[]
		\item Table \ref{table:results:MAP} and table \ref{table:result:TH} shows 
		the Intersection over Union (IoU) per class computed according to (3)
		and the averaged IoU calculated according to (4) 
		for all the proposed topologies.
		The results of topologies FCN and U1 are used as a baseline. The averaged IoU including 
		the background class is only shown for informational purposes. 
		The best results for each one of the classes have been highlighted in bold.
		
		Table \ref{table:results:MAP} shows the results obtained by using the
		MAP criterion to label each pixel in one of the target classes.
		Each pixel in the output is assigned to the class with the highest score generated by 
		the \emph{softmax} activation (see subsection 4.3). 
		
		Table \ref{table:result:TH} shows the results obtained by using the
		TH criterion to label each pixel in one of the target classes. 
		A threshold per target class was tuned using the validation subset to compute 
		the value of the IoU metric for different thresholds (see subsection 4.3). 
		
		%
		\item Topology UMD obtained the best results of all the variants tested in this work, 
		outperforming the baseline architecture U-Net (U1) for all classes using the two labelling criteria.  

		\item The second best performing topology was UAMD, slightly below UMD, 
		but slightly improves in the \emph{Vertebrae} ($IoU_{1}$),
		\emph{Sacrum} ($IoU_{2}$) and \emph{Intervertebral Disc} ($IoU_{3}$) classes.

		\item \emph{Nerve Root} ($IoU_{9}$), \emph{Blood Vessels} ($IoU_{10}$), \emph{Epidural Fat} ($IoU_{6}$) and \emph{Intramuscular Fat} ($IoU_{7}$) are the most challenging classes to be detected.

		\item Seven of the proposed topologies (UD, UAD, UMD, UAMD, UDD,  UMDD, UDD2) outperforms the two baseline architectures: 
		the standard U-Net and the FCN. 
		Four of these topologies use multi-kernels at the input and 
		Deep Supervision at the output generated by the last level of the decoder branch,
		or DS.v3.

		\item In all the results obtained performed, the TH criterion performed better than MAP criterion.

	\end{itemize}
	
	\section{Ensembles - Model Averaging approach}
	
	\begin{table*}[h!]
		\begin{center}
			\caption{Performance of the Automatic Semantic Segmentation generated by several ensembles using 
				the Model Averaging approach, computed by using the Arithmetic mean (1) (Arith).
				The Intersection over Union (IoU) was the metric used to evaluate the performance of 
				the 12 classes by using equation (3). 
				The average with and without the background class was computed equation (4). 
				-- Background is not a target class. 
				The MAP criterion 
				was used to label each single pixel into one of the target classes.} 
				\begin{tabular}{|c|l|c|c|c|c|c|c|c|c|c|c|c|}
					\hline
					\multicolumn{2}{|c|}{\textbf{Class}} & \multicolumn{10}{c|}{\textbf{Model Averaging - Ensembles according to Table 5. 
							+ Arith + MAP}} \\
					\cline{1-12}
					\textbf{\#} & \textbf{Id} & \textbf{E4} & \textbf{E5} & \textbf{E6} & \textbf{E7} & \textbf{E8} & \textbf{E9} & \textbf{E10} & \textbf{E11} & \textbf{E12} & \textbf{E13}\\
					\hline
					0 & \textbf{Background}                & $92.5\%$ & $92.5\%$ & $92.5\%$ & $92.6\%$ & $92.6\%$ & $92.6\%$ & $92.6\%$ & $92.6\%$ & $92.6\%$ & $\boldsymbol{92.6\%}$ \\
					1 & \textbf{Vert}                      & $86.7\%$ & $86.8\%$ & $86.8\%$ & $86.8\%$ & $86.8\%$ & $86.8\%$ & $86.6\%$ & $86.8\%$ & $\boldsymbol{86.8\%}$ & $86.8\%$ \\
					2 & \textbf{Sacrum}                    & $85.1\%$ & $85.2\%$ & $85.1\%$ & $85.2\%$ & $85.2\%$ & $85.2\%$ & $85.2\%$ & $85.2\%$ & $85.2\%$ & $\boldsymbol{85.2\%}$ \\
					3 & \textbf{Int-Disc}                  & $89.2\%$ & $89.3\%$ & $89.3\%$ & $89.3\%$ & $89.3\%$ & $89.3\%$ & $\boldsymbol{89.4\%}$ & $89.3\%$ & $89.36\%$ & $89.4\%$ \\
					4 & \textbf{Spinal-Cavity}             & $76.6\%$ & $76.6\%$ & $76.7\%$ & $76.8\%$ & $76.8\%$ & $76.8\%$ & $76.8\%$ & $76.8\%$ & $76.8\%$ & $\boldsymbol{76.8\%}$ \\
					5 & \textbf{SCT}                       & $92.9\%$ & $93.0\%$ & $93.0\%$ & $93.0\%$ & $93.0\%$ & $92.9\%$ & $93.0\%$ & $93.0\%$ & $\boldsymbol{93.0\%}$ & $93.0\%$ \\
					6 & \textbf{Epi-Fat}                   & $59.7\%$ & $59.7\%$ & $59.8\%$ & $59.9\%$ & $59.9\%$ & $60.0\%$ & $60.0\%$ & $60.0\%$ & $60.0\%$ & $\boldsymbol{60.0\%}$ \\
					7 & \textbf{IM-Fat}                    & $65.2\%$ & $65.3\%$ & $65.4\%$ & $65.4\%$ & $65.4\%$ & $65.5\%$ & $65.5\%$ & $65.5\%$ & $\boldsymbol{65.5\%}$ & $65.5\%$ \\
					8 & \textbf{Rper-Fat}                  & $71.8\%$ & $71.8\%$ & $71.9\%$ & $71.9\%$ & $72.0\%$ & $72.0\%$ & $72.0\%$ & $72.0\%$ & $72.0\%$ & $\boldsymbol{72.0\%}$ \\
					9 & \textbf{Nerve-Root}                & $52.8\%$ & $52.8\%$ & $52.9\%$ & $53.0\%$ & $53.1\%$ & $53.0\%$ & $53.1\%$ & $53.0\%$ & $53.1\%$ & $\boldsymbol{53.1\%}$ \\
					10 & \textbf{Blood-Vessels}            & $62.8\%$ & $62.6\%$ & $62.8\%$ & $62.8\%$ & $63.0\%$ & $63.0\%$ & $63.0\%$ & $63.0\%$ & $63.0\%$ & $\boldsymbol{63.0\%}$ \\
					11 & \textbf{Muscle}                   & $81.7\%$ & $81.7\%$ & $81.8\%$ & $81.8\%$ & $81.8\%$ & $81.9\%$ & $81.9\%$ & $81.9\%$ & $81.9\%$ & $\boldsymbol{81.9\%}$ \\
					\hline
					\multicolumn{2}{|l|}{\textbf{IoU} without Bg.}      & $75.0\%$ & $75.0\%$ & $75.0\%$ & $75.1\%$ & $75.1\%$ & $75.1\%$ & $75.1\%$ & $75.1\%$ & $75.2\%$ & $\boldsymbol{75.2\%}$ \\
					\multicolumn{2}{|l|}{\textbf{IoU} with Bg.}        & $76.4\%$ & $76.5\%$ & $76.5\%$ & $76.5\%$ & $76.6\%$ & $76.6\%$ & $76.6\%$ & $76.6\%$ & $76.6\%$ & $\boldsymbol{76.6\%}$ \\
					\hline
				\end{tabular}
			\label{table:results:AA-MAP}
		\end{center}
	\end{table*}
	
	\begin{table*}[h!]
		\begin{center}
			
			\caption{Performance of the Automatic Semantic Segmentation generated by several ensembles using 
				the Model Averaging	approach, computed by using the Geometric mean (2) (Geo).
				The Intersection over Union (IoU) was the metric used to evaluate 
				the performance of the 12 classes by using equation (3). 
				The average with and without the background class was computed equation (4). 
				-- Background is not a target class. 
				The MAP criterion 
				was used to label each single pixel into one of the target classes.} 
			
				\begin{tabular}{|c|l|c|c|c|c|c|c|c|c|c|c|c|}
					\hline
					\multicolumn{2}{|c|}{\textbf{Class}} & \multicolumn{10}{c|}{\textbf{Model Averaging - Ensembles according to Table 5. 
								+ Geo + MAP}} \\
						\cline{1-12}
						\textbf{\#} & \textbf{Id} & \textbf{E4} & \textbf{E5} & \textbf{E6} & \textbf{E7} & \textbf{E8} & \textbf{E9} & \textbf{E10} & \textbf{E11} & \textbf{E12} & \textbf{E13}\\
						\hline
						0 & \textbf{Background}                & $92,5\%$ & $92,5\%$ & $92,5\%$ & $92,6\%$ & $92,6\%$ & $92,6\%$ & $92,6\%$ & $92,6\%$ & $92,6\%$ & $\boldsymbol{92,6\%}$ \\
						1 & \textbf{Vert}                      & $86,7\%$ & $86,8\%$ & $86,8\%$ & $86,9\%$ & $86,8\%$ & $86,9\%$ & $86,9\%$ & $86,9\%$ & $\boldsymbol{86,9\%}$ & $86,9\%$ \\
						2 & \textbf{Sacrum}                    & $85,1\%$ & $85,2\%$ & $85,2\%$ & $85,2\%$ & $85,2\%$ & $85,2\%$ & $85,3\%$ & $85,2\%$ & $85,3\%$ & $\boldsymbol{85,3\%}$ \\
						3 & \textbf{Int-Disc}                  & $89,2\%$ & $89,3\%$ & $89,3\%$ & $89,4\%$ & $89,3\%$ & $89,4\%$ & $\boldsymbol{89,4\%}$ & $89,4\%$ & $89,4\%$ & $89,4\%$ \\
						4 & \textbf{Spinal-Cavity}             & $76,6\%$ & $76,6\%$ & $76,7\%$ & $76,7\%$ & $76,7\%$ & $76,7\%$ & $76,8\%$ & $76,8\%$ & $76,8\%$ & $\boldsymbol{76,8\%}$ \\
						5 & \textbf{SCT}                       & $92,9\%$ & $93,0\%$ & $93,0\%$ & $93,0\%$ & $93,0\%$ & $93,0\%$ & $93,0\%$ & $93,0\%$ & $\boldsymbol{93,0\%}$ & $93,0\%$ \\
						6 & \textbf{Epi-Fat}                   & $59,7\%$ & $59,7\%$ & $59,9\%$ & $59,9\%$ & $59,9\%$ & $60,0\%$ & $60,0\%$ & $60,0\%$ & $60,0\%$ & $\boldsymbol{60,0\%}$ \\
						7 & \textbf{IM-Fat}                    & $65,2\%$ & $65,3\%$ & $65,4\%$ & $65,4\%$ & $65,4\%$ & $65,5\%$ & $65,5\%$ & $65,5\%$ & $\boldsymbol{65,5\%}$ & $65,5\%$ \\
						8 & \textbf{Rper-Fat}                  & $71,8\%$ & $71,8\%$ & $71,9\%$ & $71,9\%$ & $72,0\%$ & $72,0\%$ & $72,0\%$ & $72,0\%$ & $72,0\%$ & $\boldsymbol{72,0\%}$ \\
						9 & \textbf{Nerve-Root}                & $52,8\%$ & $52,9\%$ & $52,9\%$ & $53,0\%$ & $53,0\%$ & $53,0\%$ & $53,1\%$ & $53,0\%$ & $53,1\%$ & $\boldsymbol{53,1\%}$ \\
						10 & \textbf{Blood-Vessels}            & $62,8\%$ & $62,6\%$ & $62,8\%$ & $62,8\%$ & $62,8\%$ & $63,0\%$ & $63,0\%$ & $63,0\%$ & $63,0\%$ & $\boldsymbol{63,0\%}$ \\
						11 & \textbf{Muscle}                   & $81,8\%$ & $81,8\%$ & $81,8\%$ & $81,8\%$ & $81,8\%$ & $81,9\%$ & $81,9\%$ & $81,9\%$ & $81,9\%$ & $\boldsymbol{81,9\%}$ \\
						\hline
						\multicolumn{2}{|l|}{\textbf{IoU} without Bg.}      & $75,0\%$ & $75,0\%$ & $75,0\%$ & $75,1\%$ & $75,1\%$ & $75,1\%$ & $75,2\%$ & $75,1\%$ & $75,2\%$ & $\boldsymbol{75,2\%}$ \\
						\multicolumn{2}{|l|}{\textbf{IoU} with Bg.}        & $76,4\%$ & $76,5\%$ & $76,5\%$ & $76,5\%$ & $76,6\%$ & $76,6\%$ & $76,6\%$ & $76,6\%$ & $76,6\%$ & $\boldsymbol{76,6\%}$ \\
						\hline
					\end{tabular}
				\label{table:results:GA-MAP}
			\end{center}
		\end{table*}
		For this experiment, the output of several networks corresponding to different topologies 
		is combined to form a classifier that is an ensemble of classifiers using 
		the model averaging approach (see Figure 5).
		
		Model averaging is a technique where $R$ models equally contribute to obtaining the ensemble's output. 
		Two ways of computing the output of the ensemble from the output of the components were considered, 
		the arithmetic mean (Arith)  and the geometric mean (Geo) (see Subsection 4.2.1.).
		
		In addition to training and evaluating individual semantic segmentation models designed as 
		variations from the U-Net architecture (see \ref{sect:Experiment:1}),
		a set of ensembles were created by grouping from 4 to 13 models.
		Table 5 show all the ensembles used in this work, where it can be observed that the FCN network 
		was only used in ensembles E8 and E13 for comparison purposes.
		
		The experiments for each evaluated network topology or ensemble were carried out following 
		the same 3-fold cross-validation procedure (See Subsection 3.1.3.).
		
		The proposed network topologies use the \emph{softmax} activation function in the output layer; 
		their outputs are Normalised and sum 1. We refer to them as vectors of normalised scores.
		
		The models output masks corresponding to $256 \times 256$ patches are combined and generate a 
		single mask per original slide (medical image) to evaluate the quality of 
		the automatic semantic segmentation.

		\medskip
		The output of the ensemble is also one vector of normalised scores per pixel. 
		The reported results were computed after labelling each single pixel to one of 
		the 12 classes by either the MAP criterion (see Subsection 4.3).
		

		\subsection{Results}
		
		\begin{itemize}[]
			\item Table \ref{table:results:AA-MAP} and  Table \ref{table:results:GA-MAP} shows 
			the Intersection over Union (IoU) per class computed according to (3)
			and the averaged IoU calculated according to (4) 
			for all ensembles using the \textit{Model Averaging} approach,
			The averaged IoU including the background class is only shown for informational purposes.
			Two ways of computing the output of the ensemble from the output of the components were used:
			\textit{Arithmetic mean} (1) (Arith) and \textit{Geometric mean} (2) (Geo),
			the results shown in \ref{table:results:AA-MAP} and Table \ref{table:results:GA-MAP} respectively. 
			\emph{Maximum a Posteriori Probability Estimation} (MAP) criteria described in Subsection 4.3 
			were used to label each single pixel into one of the	target classes. 
			The best results for each one of the classes have been highlighted in bold.
			
			\item Ensemble \textit{E}13 obtained the best results of all the ensembles. 

			\item  No significant differences between the arithmetic mean and the geometric mean are observed.
			The high similarity between both ways of computing the mean confirms 
			that all the topologies combined in the ensembles perform very similarly.

			\item Compared to table \ref{table:results:MAP} in experiment  \ref{sect:Experiment:1} , 
			all ensemble results calculated with the model averaging approach outperforming 
			the baseline architecture U-Net (U1) and the best performing architecture (UMD) in all classes.

		\end{itemize}

		\section{Ensembles - Stacking Model approach}
		
		For this experiment, the stacking model tries to learn to obtain the better combination
		of the predictions of $R$ single models in order to obtain the best prediction.
		
		An ensemble following the stacking model is implemented in three stages: 
		(a) \emph{layer merging}, 
		(b) \emph{meta-learner}, and 
		(c) \emph{prediction},
		As indicated in Subsection 4.2.1.
		
		%
		The stacking model technique was used with two different approaches
		to prepare the input to the layer-merging stage:
		(a) the output of the \emph{softmax} activation layer from each model $r$ in the ensemble, i.e., the vector $y_r$, and
		(b) the 64-channel tensor at the input to the classification block, i.e., 
		the output generated by the last level of the decoder branch, 
		or the last level of the deep supervision block (DS.v3) when applicable.
		The combination of the inputs in the layer-merging stage can be done by concatenation, averaging, or adding.
		
		The set of ensembles created by grouping 4 to 13 models is shown in table 5.
		
		Ensembles based on the stacking model were trained during 50 epochs using the same 
		data-augmentation transformations used to train each single network (see Subsection 5.1), and 
		following the 3-fold cross-validation procedure with the same partitions of the dataset.

		Table 6 shows the input formats and layer configurations for 
		the best performing ensembles based on the stacking model assembling technique. 
		Ensemble configurations are identified by a three letter acronynm. 
		First letter identifies the input type, 
		\textbf{N} and \textbf{T} which stand for normalised scores (\emph{softmax} output) and 64-channels tensors, respectively. 
		Second letter indicates layer merging operator: averaging (\textbf{A}) and concatenation (\textbf{C}). 
		The addition operator was also used in the whole experimentation. However, 
		its results are not presented here because they were too poor.
		Third and last letter corresponds to the type of meta-learner used, in this case only dense layers were used, 
		so the third letter is fixed to \textbf{D}.

		The stacking model output masks corresponding to $256 \times 256$ patches are used to be combined and 
		generate a single mask per original slide (medical image) to evaluate the quality of the automatic semantic segmentation.
		The vector corresponding to every single pixel of the reconstructed mask is used to assign each pixel to one of the 12 classes 
		by either the MAP criterion or the TH strategy (see Subsection 4.3).
		
		
		\subsection{Results}
		
		\begin{itemize}[]
			%
			\item Table 6 shows the best combination of layers obtained for stacking models based on the input type. 
			We get two stacking model configuration options, NAD and TCD.

			In NAD configuration, the inputs to the stacking model are \emph{normalised scores}, the layer-merging is Average, and the meta-learner is a dense layer. 

			In TCD configuration, the inputs to the stacking model are 64-channel \emph{tensors} at the input of the classification block from each model. The merging layer is a concatenated layer, and the meta-learner is a dense layer.

			\item Tables \ref{table:results:NAD:TH}, \ref{table:results:NAD:MAP},
			\ref{table:results:TCD:TH} and \ref{table:results:TCD:MAP}
			shows the Intersection over Union (IoU) per class computed according to (3)
			and the averaged IoU calculated according to (4) 
			for all ensembles using the stacking model approach,
			The averaged IoU including the background class is only shown for informational purposes.
			
			The results obtained with the two stacking model configurations and the respective method used to label each pixel in one of the target classes are shown in the tables as follows:	
			NAD + TH in Table \ref{table:results:NAD:TH}, 
			NAD + MAP in Table \ref{table:results:NAD:MAP}, 
			TCD + TH in Table \ref{table:results:TCD:TH} and 
			TCD + MAP in Table \ref{table:results:TCD:MAP}. 
			The best results for each one of the classes have been highlighted in bold.
			\item The ensemble \textit{E}12+NAD+TH obtained the best overall results.
			Let us remark that the TH labelling criterion performed better than 
			the MAP criterion in all the performed experiments. 
			
			\item The ensembles \textit{E}10+TCD+TH and \textit{E}12+NAD+TH performed significantly better than 
			best performing topology tested in this work (UMD+TH) for all target classes.

			\item The ensembles including the FCN topology, E8 and E13, have a significant performance drop. 
			Comparing \textit{E}12 and \textit{E}13 results for the configuration NAD+TH, 
			it can be observed that the addition of 
			the FCN topology significantly deteriorates the performance.

			\item It can be observed that the model averaging assembling technique is more robust than 
			the stacking model technique to the variance resulting from the predictions of 
			the networks that constitute the ensemble.
			
		\end{itemize}
		\begin{table*}[t!]
			\begin{center}
				
				\caption{Performance of the Automatic Semantic Segmentation generated by several stacking model ensembles with NAD configuration. 
					measured in terms of the Intersection over Union (IoU) metric on 12 classes.
					for each single class it was used the formula from (3)
					the average with and without the background class was computed using the formula from (4)
					-- Background is not a target class. 
					The Threshold Optimisation (TH) criteria described in Subsection 4.3
					was used to label each single pixel into one of the target classes.}
					\begin{tabular}{|c|l|c|c|c|c|c|c|c|c|c|c|c|}
						\hline
						\multicolumn{2}{|c|}{\textbf{Class}} & \multicolumn{10}{c|}{\textbf{Stacking Model - Ensembles according to Table 5. 
									+ NAD + TH}} \\
							\cline{1-12}
							\textbf{\#} & \textbf{Id} & \textbf{E4} & \textbf{E5} & \textbf{E6} & \textbf{E7} & \textbf{E8} & \textbf{E9} & \textbf{E10} & \textbf{E11} & \textbf{E12} & \textbf{E13}\\
							\hline
							0 & \textbf{Background}                & $92.5\%$ & $92.5\%$ & $92.6\%$ & $92.6\%$ & $92.3\%$ & $92.6\%$ & $92.6\%$ & $\boldsymbol{92.6\%}$ & $92.6\%$ & $92.6\%$\\
							1 & \textbf{Vert}                      & $86.9\%$ & $86.9\%$ & $87.0\%$ & $87.0\%$ & $86.9\%$ & $87.0\%$ & $\boldsymbol{87.0\%}$ & $87.0\%$ & $87.0\%$ & $86.9\%$\\
							2 & \textbf{Sacrum}                    & $85.1\%$ & $85.1\%$ & $85.2\%$ & $85.2\%$ & $85.2\%$ & $85.3\%$ & $85.3\%$ & $85.4\%$ & $\boldsymbol{85.4\%}$ & $85.2\%$\\
							3 & \textbf{Int-Disc}                  & $89.4\%$ & $89.4\%$ & $89.4\%$ & $89.5\%$ & $89.4\%$ & $89.5\%$ & $89.4\%$ & $\boldsymbol{89.5\%}$ & $89.5\%$ & $89.4\%$\\
							4 & \textbf{Spinal-Cavity}             & $76.7\%$ & $76.7\%$ & $76.8\%$ & $76.8\%$ & $76.3\%$ & $76.9\%$ & $76.7\%$ & $76.8\%$ & $\boldsymbol{77.0\%}$ & $76.7\%$\\
							5 & \textbf{SCT}                       & $92.9\%$ & $93.0\%$ & $93.0\%$ & $93.0\%$ & $93.0\%$ & $93.0\%$ & $93.1\%$ & $93.0\%$ & $93.1\%$ & $\boldsymbol{93.1\%}$\\
							6 & \textbf{Epi-Fat}                   & $59.6\%$ & $59.6\%$ & $59.8\%$ & $59.8\%$ & $58.4\%$ & $59.9\%$ & $\boldsymbol{60.0\%}$ & $59.9\%$ & $60.0\%$ & $59.8\%$\\
							7 & \textbf{IM-Fat}                    & $65.4\%$ & $65.5\%$ & $65.5\%$ & $65.6\%$ & $65.2\%$ & $65.6\%$ & $65.7\%$ & $65.7\%$ & $\boldsymbol{65.7\%}$ & $65.6\%$\\
							8 & \textbf{Rper-Fat}                  & $71.6\%$ & $71.8\%$ & $71.9\%$ & $71.9\%$ & $\boldsymbol{72.3\%}$ & $72.0\%$ & $72.0\%$ & $72.0\%$ & $72.0\%$ & $72.0\%$\\
							9 & \textbf{Nerve-Root}                & $53.0\%$ & $53.0\%$ & $53.3\%$ & $53.2\%$ & $52.2\%$ & $53.3\%$ & $\boldsymbol{53.3\%}$ & $53.3\%$ & $53.3\%$ & $52.2\%$\\
							10 & \textbf{Blood-Vessels}            & $62.8\%$ & $62.9\%$ & $63.0\%$ & $63.0\%$ & $63.3\%$ & $63.2\%$ & $63.2\%$ & $\boldsymbol{63.4\%}$ & $63.3\%$ & $63.3\%$\\
							11 & \textbf{Muscle}                   & $81.8\%$ & $81.9\%$ & $81.9\%$ & $82.0\%$ & $81.8\%$ & $82.1\%$ & $82.0\%$ & $\boldsymbol{82.1\%}$ & $82.0\%$ & $82.0\%$\\
							\hline
							\multicolumn{2}{|l|}{\textbf{IoU} without Bg.}     & $75.0\%$ & $75.1\%$ & $75.2\%$ & $75.2\%$ & $74.9\%$ & $75.2\%$ & $75.2\%$ & $75.3\%$ & $\boldsymbol{75.3\%}$ & $75.1\%$ \\
							\multicolumn{2}{|l|}{\textbf{IoU} with Bg.}        & $76.5\%$ & $76.5\%$ & $76.6\%$ & $76.6\%$ & $76.3\%$ & $76.7\%$ & $76.7\%$ & $76.7\%$ & $\boldsymbol{76.7\%}$ & $76.6\%$ \\
							\hline
						\end{tabular}
					\label{table:results:NAD:TH}
				\end{center}
			\end{table*}
			
			\begin{table*}[t!]
				\begin{center}
					
					\caption{Performance of the Automatic Semantic Segmentation generated by several stacking model ensembles with NAD configuration. 
						measured in terms of the Intersection over Union (IoU) metric on 12 classes.
						for each single class it was used the formula from (3)
						the average with and without the background class was computed using the formula from (4)
						-- Background is not a target class. 
						The Maximum a Posteriori Probability Estimate (MAP) criteria described in Subsection 4.3
						was used to label each single pixel into one of the target classes.}
						\begin{tabular}{|c|l|c|c|c|c|c|c|c|c|c|c|c|}
							\hline
							\multicolumn{2}{|c|}{\textbf{Class}} & \multicolumn{10}{c|}{\textbf{Stacking Model - Ensembles according to Table 5. 
										+ NAD + MAP}} \\
								\cline{1-12}
								\textbf{\#} & \textbf{Id} & \textbf{E4} & \textbf{E5} & \textbf{E6} & \textbf{E7} & \textbf{E8} & \textbf{E9} & \textbf{E10} & \textbf{E11} & \textbf{E12} & \textbf{E13}\\
								\hline
								0 & \textbf{Background}                & $92.4\%$ & $92.5\%$ & $92.5\%$ & $92.6\%$ & $92.3\%$ & $92.6\%$ & $92.6\%$ & $\boldsymbol{92.6\%}$ & $92.5\%$ & $92.5\%$ \\
								1 & \textbf{Vert}                      & $86.7\%$ & $86.9\%$ & $87.0\%$ & $86.7\%$ & $86.8\%$ & $\boldsymbol{87.0\%}$ & $86.9\%$ & $86.9\%$ & $86.9\%$ & $86.9\%$ \\
								2 & \textbf{Sacrum}                    & $84.9\%$ & $84.9\%$ & $84.8\%$ & $85.1\%$ & $84.9\%$ & $85.1\%$ & $84.9\%$ & $85.1\%$ & $\boldsymbol{85.1\%}$ & $85.1\%$ \\
								3 & \textbf{Int-Disc}                  & $89.2\%$ & $89.3\%$ & $89.4\%$ & $89.3\%$ & $89.3\%$ & $89.4\%$ & $89.4\%$ & $\boldsymbol{89.4\%}$ & $89.3\%$ & $89.3\%$ \\
								4 & \textbf{Spinal-Cavity}             & $76.4\%$ & $76.3\%$ & $76.5\%$ & $76.5\%$ & $75.9\%$ & $\boldsymbol{76.7\%}$ & $76.30\%$ & $76.5\%$ & $76.5\%$ & $76.2\%$ \\
								5 & \textbf{SCT}                       & $92.9\%$ & $93.0\%$ & $93.0\%$ & $93.0\%$ & $92.9\%$ & $93.0\%$ & $93.0\%$ & $93.0\%$ & $93.0\%$ & $\boldsymbol{93.0\%}$ \\
								6 & \textbf{Epi-Fat}                   & $59.3\%$ & $59.3\%$ & $59.5\%$ & $59.5\%$ & $58.0\%$ & $59.6\%$ & $\boldsymbol{59.6\%}$ & $59.6\%$ & $59.3\%$ & $59.4\%$ \\
								7 & \textbf{IM-Fat}                    & $65.2\%$ & $65.3\%$ & $65.3\%$ & $\boldsymbol{65.4\%}$ & $65.0\%$ & $65.4\%$ & $65.4\%$ & $65.4\%$ & $65.4\%$ & $65.4\%$ \\
								8 & \textbf{Rper-Fat}                  & $71.6\%$ & $71.7\%$ & $71.8\%$ & $71.9\%$ & $\boldsymbol{72.1\%}$ & $71.9\%$ & $72.0\%$ & $71.9\%$ & $71.9\%$ & $72.0\%$ \\
								9 & \textbf{Nerve-Root}                & $52.6\%$ & $52.7\%$ & $\boldsymbol{53.0\%}$ & $52.7\%$ & $51.8\%$ & $52.8\%$ & $53.0\%$ & $52.9\%$ & $51.8\%$ & $51.8\%$ \\
								10 & \textbf{Blood-Vessels}            & $62.6\%$ & $62.6\%$ & $62.8\%$ & $62.7\%$ & $63.0\%$ & $63.0\%$ & $62.9\%$ & $\boldsymbol{63.1\%}$ & $62.8\%$ & $62.9\%$ \\
								11 & \textbf{Muscle}                   & $81.7\%$ & $81.8\%$ & $81.8\%$ & $81.8\%$ & $81.7\%$ & $\boldsymbol{81.9\%}$ & $81.9\%$ & $81.9\%$ & $81.9\%$ & $81.9\%$ \\
								\hline
								\multicolumn{2}{|l|}{\textbf{IoU} without Bg.}     & $74.8\%$ & $74.9\%$ & $75.0\%$ & $75.0\%$ & $74.7\%$ & $\boldsymbol{75.1\%}$ & $75.0\%$ & $75.1\%$ & $74.9\%$ & $74.9\%$ \\
								\multicolumn{2}{|l|}{\textbf{IoU} with Bg.}        & $76.3\%$ & $76.4\%$ & $76.4\%$ & $76.4\%$ & $76.1\%$ & $\boldsymbol{76.5\%}$ & $76.5\%$ & $76.5\%$ & $76.4\%$ & $76.4\%$ \\
								\hline
							\end{tabular}
						\label{table:results:NAD:MAP}
					\end{center}
				\end{table*}
				\begin{table*}[t!]
					\begin{center}
						
						\caption{Performance of the Automatic Semantic Segmentation generated by several stacking model ensembles with TCD configuration. 
							measured in terms of the Intersection over Union (IoU) metric on 12 classes.
							for each single class it was used the formula from (3)
							the average with and without the background class was computed using the formula from (4)
							-- Background is not a target class. 
							The Threshold Optimisation (TH) criteria described in Subsection 4.3
							was used to label each single pixel into one of the target classes.}
							\begin{tabular}{|c|l|c|c|c|c|c|c|c|c|c|c|c|}
								\hline
								\multicolumn{2}{|c|}{\textbf{Class}} & \multicolumn{10}{c|}{\textbf{Stacking Model - Ensembles according to Table 5. 
											+ TCD + TH}} \\
									\cline{1-12}
									\textbf{\#} & \textbf{Id} & \textbf{E4} & \textbf{E5} & \textbf{E6} & \textbf{E7} & \textbf{E8} & \textbf{E9} & \textbf{E10} & \textbf{E11} & \textbf{E12} & \textbf{E13}\\
									\hline
									0 & \textbf{Background}                & $92.4\%$ & $92.4\%$ & $92.4\%$ & $92.4\%$ & $92.1\%$ & $92.5\%$ & $92.5\%$ & $\boldsymbol{92.4\%}$ & $92.4\%$ & $92.4\%$ \\
									1 & \textbf{Vert}                      & $86.6\%$ & $86.7\%$ & $86.7\%$ & $86.7\%$ & $86.6\%$ & $86.7\%$ & $\boldsymbol{86.7\%}$ & $86.7\%$ & $86.7\%$ & $86.7\%$ \\
									2 & \textbf{Sacrum}                    & $85.0\%$ & $85.0\%$ & $84.8\%$ & $85.0\%$ & $84.8\%$ & $84.9\%$ & $\boldsymbol{85.0\%}$ & $85.0\%$ & $85.0\%$ & $84.9\%$ \\
									3 & \textbf{Int-Disc}                  & $89.2\%$ & $89.2\%$ & $89.2\%$ & $89.3\%$ & $89.2\%$ & $89.2\%$ & $\boldsymbol{89.3\%}$ & $89.2\%$ & $89.3\%$ & $89.3\%$ \\
									4 & \textbf{Spinal-Cavity}             & $76.3\%$ & $76.3\%$ & $76.4\%$ & $76.3\%$ & $75.7\%$ & $76.3\%$ & $\boldsymbol{76.5\%}$ & $76.4\%$ & $76.3\%$ & $76.3\%$ \\
									5 & \textbf{SCT}                       & $92.8\%$ & $92.8\%$ & $92.8\%$ & $92.8\%$ & $92.8\%$ & $92.9\%$ & $92.9\%$ & $92.9\%$ & $\boldsymbol{92.9\%}$ & $92.9\%$ \\
									6 & \textbf{Epi-Fat}                   & $59.2\%$ & $59.3\%$ & $59.2\%$ & $59.2\%$ & $57.9\%$ & $59.4\%$ & $\boldsymbol{59.4\%}$ & $59.4\%$ & $59.2\%$ & $59.2\%$ \\
									7 & \textbf{IM-Fat}                    & $64.8\%$ & $64.9\%$ & $64.8\%$ & $64.9\%$ & $64.4\%$ & $65.0\%$ & $\boldsymbol{65.1\%}$ & $65.0\%$ & $65.0\%$ & $65.0\%$ \\
									8 & \textbf{Rper-Fat}                  & $71.3\%$ & $71.5\%$ & $71.5\%$ & $71.5\%$ & $\boldsymbol{71.7\%}$ & $71.6\%$ & $71.6\%$ & $71.6\%$ & $71.6\%$ & $71.6\%$ \\
									9 & \textbf{Nerve-Root}                & $52.5\%$ & $52.5\%$ & $52.3\%$ & $52.4\%$ & $51.2\%$ & $52.3\%$ & $\boldsymbol{52.6\%}$ & $52.3\%$ & $51.5\%$ & $51.5\%$ \\
									10 & \textbf{Blood-Vessels}            & $62.51\%$ & $62.3\%$ & $62.3\%$ & $62.2\%$ & $62.2\%$ & $\boldsymbol{62.7\%}$ & $62.6\%$ & $62.6\%$ & $62.7\%$ & $62.7\%$ \\
									11 & \textbf{Muscle}                   & $81.5\%$ & $81.4\%$ & $81.5\%$ & $81.5\%$ & $81.3\%$ & $81.6\%$ & $81.6\%$ & $81.6\%$ & $81.6\%$ & $\boldsymbol{81.7\%}$ \\
									
									\hline
									\multicolumn{2}{|l|}{\textbf{IoU} without Bg.}     & $74.7\%$ & $74.7\%$ & $74.7\%$ & $74.7\%$ & $74.3\%$ & $74.8\%$ & $\boldsymbol{74.8\%}$ & $74.8\%$ & $74.7\%$ & $74.7\%$ \\
									\multicolumn{2}{|l|}{\textbf{IoU} with Bg.}        & $76.2\%$ & $76.2\%$ & $76.2\%$ & $76.2\%$ & $75.8\%$ & $76.2\%$ & $\boldsymbol{76.3\%}$ & $76.3\%$ & $76.2\%$ & $76.2\%$ \\
									\hline
								\end{tabular}
							\label{table:results:TCD:TH}
						\end{center}
					\end{table*}

					\begin{table*}[t!]
						\begin{center}
							
							\caption{Performance of the Automatic Semantic Segmentation generated by several stacking model ensembles with TCD configuration. 
								measured in terms of the Intersection over Union (IoU) metric on 12 classes.
								for each single class it was used the formula from (3)
								the average with and without the background class was computed using the formula from (4)
								-- Background is not a target class. 
								The Maximum a Posteriori Probability Estimate (MAP) criteria described in Subsection 4.3
								was used to label each single pixel into one of the target classes.}

							\begin{tabular}{|c|l|c|c|c|c|c|c|c|c|c|c|c|}
								\hline
								\multicolumn{2}{|c|}{\textbf{Class}} & \multicolumn{10}{c|}{\textbf{Stacking Model - Ensembles according to Table 5. 
											+ TCD + MAP}} \\
									\cline{1-12}
									\textbf{\#} & \textbf{Id} & \textbf{E4} & \textbf{E5} & \textbf{E6} & \textbf{E7} & \textbf{E8} & \textbf{E9} & \textbf{E10} & \textbf{E11} & \textbf{E12} & \textbf{E13}\\
									\hline
									0 & \textbf{Background}                & $92.3\%$ & $92.4\%$ & $92.4\%$ & $92.4\%$ & $92.0\%$ & $\boldsymbol{92.4\%}$ & $92.4\%$ & $92.4\%$ & $92.4\%$ & $92.4\%$ \\
									1 & \textbf{Vert}                      & $86.4\%$ & $86.5\%$ & $86.5\%$ & $86.6\%$ & $86.4\%$ & $86.5\%$ & $\boldsymbol{86.6\%}$ & $86.6\%$ & $86.6\%$ & $86.5\%$ \\
									2 & \textbf{Sacrum}                    & $84.8\%$ & $\boldsymbol{84.8\%}$ & $84.8\%$ & $84.8\%$ & $84.6\%$ & $84.8\%$ & $84.8\%$ & $84.8\%$ & $84.8\%$ & $84.8\%$ \\
									3 & \textbf{Int-Disc}                  & $89.0\%$ & $89.1\%$ & $89.0\%$ & $89.1\%$ & $89.1\%$ & $89.1\%$ & $89.1\%$ & $89.1\%$ & $\boldsymbol{89.1\%}$ & $89.1\%$ \\
									4 & \textbf{Spinal-Cavity}             & $75.9\%$ & $76.0\%$ & $76.0\%$ & $76.0\%$ & $75.5\%$ & $76.1\%$ & $76.1\%$ & $\boldsymbol{76.1\%}$ & $76.0\%$ & $76.0\%$ \\
									5 & \textbf{SCT}                       & $92.7\%$ & $92.8\%$ & $92.7\%$ & $92.8\%$ & $92.7\%$ & $92.8\%$ & $\boldsymbol{92.8\%}$ & $92.8\%$ & $92.8\%$ & $92.8\%$ \\
									6 & \textbf{Epi-Fat}                   & $58.8\%$ & $59.0\%$ & $58.8\%$ & $58.9\%$ & $57.6\%$ & $58.2\%$ & $\boldsymbol{59.1\%}$ & $58.9\%$ & $58.9\%$ & $58.9\%$ \\
									7 & \textbf{IM-Fat}                    & $64.6\%$ & $64.6\%$ & $64.6\%$ & $64.6\%$ & $64.2\%$ & $64.7\%$ & $\boldsymbol{64.8\%}$ & $64.7\%$ & $64.7\%$ & $64.7\%$ \\
									8 & \textbf{Rper-Fat}                  & $71.3\%$ & $71.4\%$ & $71.4\%$ & $71.4\%$ & $\boldsymbol{71.6\%}$ & $71.50\%$ & $71.6\%$ & $71.5\%$ & $71.5\%$ & $71.5\%$ \\
									9 & \textbf{Nerve-Root}                & $52.0\%$ & $51.9\%$ & $51.8\%$ & $51.8\%$ & $50.7\%$ & $51.7\%$ & $\boldsymbol{52.0\%}$ & $51.6\%$ & $50.9\%$ & $51.0\%$ \\
									10 & \textbf{Blood-Vessels}            & $62.2\%$ & $62.0\%$ & $61.9\%$ & $61.9\%$ & $61.9\%$ & $62.3\%$ & $62.3\%$ & $62.2\%$ & $\boldsymbol{62.3\%}$ & $62.3\%$ \\
									11 & \textbf{Muscle}                   & $81.3\%$ & $81.3\%$ & $81.3\%$ & $81.3\%$ & $81.1\%$ & $81.4\%$ & $81.4\%$ & $81.4\%$ & $81.4\%$ & $\boldsymbol{81.5\%}$ \\
									\hline
									\multicolumn{2}{|l|}{\textbf{IoU} without Bg.}     & $74.5\%$ & $74.5\%$ & $74.4\%$ & $74.5\%$ & $74.1\%$ & $74.5\%$ & $\boldsymbol{74.6\%}$ & $74.5\%$ & $74.5\%$ & $74.5\%$ \\
									\multicolumn{2}{|l|}{\textbf{IoU} with Bg.}        & $76.0\%$ & $76.0\%$ & $75.9\%$ & $75.7\%$ & $75.6\%$ & $76.0\%$ & $\boldsymbol{76.1\%}$ & $76.0\%$ & $76.0\%$ & $75.9\%$ \\
									\hline
								\end{tabular}
								
								\label{table:results:TCD:MAP}
							\end{center}
						\end{table*}
						
						\bibliography{mybibfile}